\documentclass[tightenlines,superscriptaddress,floatfixolumn,preprint,footinbib]{revtex4-1} 
\usepackage{graphicx}
\usepackage{amsmath}
\usepackage{amsfonts,amsbsy}
\usepackage{amssymb}

\def\Npb{N_{\rm  part}^{\rm Pb}}
\def\Npp{N_{\rm part}^{\rm proton}}

\def\ptasc{p^{\textrm{asc}}_T}
\def\ptrig{p^{\textrm{trig}}_T}
\def\ptmin{p^{\textrm{min}}_T}
\def\Ntrk{N_{\rm trk}^{\rm offline}}

\def\Ntrig{N_{\rm trig}}

\def\tN{\tilde{N}_{\rm trig}}
\def\as{\alpha_S}
\def\sp{S_\perp}
\def\qs{Q_S}
\def\Qs{\qs}
\def\nc{N_c}
\def\cf{C_F}
 
\def\qp{ {\bf q}_T } 
 
\def\pp{ {\bf p}_T } 
\def\kp{ {\bf k}_T } 
 
\newcommand{\kpn}[1]{ {\bf k}_{#1\perp} } 
\newcommand{\qpn}[1]{ {\bf q}_{#1\perp} }
\newcommand{\ud}{\mathrm{d}}

\begin{document}

\title{Comparison of the Color Glass Condensate to di-hadron correlations in proton-proton and 
proton-nucleus collisions}

\author{Kevin Dusling}
\affiliation{Physics Department, North Carolina State University, Raleigh, NC 27695, USA
}
\author{Raju Venugopalan}
\affiliation{Physics Department, Brookhaven National Laboratory,
  Upton, NY 11973, USA
}

\begin{abstract}
We perform a detailed comparison of long range rapidity correlations in the Color Glass Condensate framework to high multiplicity di-hadron data in proton-proton and proton-lead collisions from the CMS, ALICE and ATLAS experiments at the LHC. 
The overall good agreement thus far of the non-trivial systematics of theory
with data is strongly suggestive of gluon saturation and the presence of subtle
quantum interference effects between rapidity separated gluons. In particular,
the yield of pairs collimated in their relative azimuthal angle $\Delta
\phi\sim 0$, is sensitive to the shape of unintegrated gluon distributions in
the hadrons that are renormalization group evolved in rapidity from the beam
rapidities to those of the measured hadrons.  We present estimates for the collimated di-hadron yield expected in central deuteron-gold collisions at RHIC. 
\end{abstract}

\maketitle

\section{Introduction}

In two recent papers~\cite{Dusling:2012cg,Dusling:2012wy}, we argued that data on two particle correlations in high multiplicity proton-proton and proton-nucleus collisions from the CMS collaboration~\cite{Khachatryan:2010gv,CMS:2012qk} provided strong evidence for gluon saturation and the Color Glass Condensate (CGC) effective field theory (EFT)~\cite{Gelis:2010nm} describing this phenomenon.  These correlations, which are long range in the relative rapidity $\Delta \eta$ between pairs of charged hadrons, show an unexpected ``nearside" collimation  $\Delta \phi \approx 0$ in their relative azimuthal angle. This unusual collimation is called the ``ridge" due to its structure in the $\Delta\eta-\Delta\phi$ plane. Reviews of this ridge effect can be found in  ~\cite{Li:2012hc,Kovner:2012jm}. 

In the CGC effective theory, the nearside collimation is obtained from QCD graphs called ``Glasma graphs"; for high occupancy gluons (with transverse momenta $k_\perp \leq Q_S$, the saturation scale), Glasma graphs are enhanced by $\alpha_S^{-8}$, a factor of $\sim 10^5$ for typical values of the probed QCD fine structure constant $\alpha_S$. In the power counting of the EFT, the effect of gluon saturation on Glasma graphs ensures they provide a significant  additional contribution in high multiplicity events to ``di-jet" QCD graphs. The latter are kinematically constrained to provide an ``awayside" back-to-back collimation peaked at $\Delta \phi\approx \pi$ but do not provide a significant nearside collimation.

The importance of Glasma graphs was first discussed in ~\cite{Dumitru:2008wn} and the formalism developed in ~\cite{Gelis:2008sz,Dusling:2009ni}. It was first postulated as an explanation of the high multiplicity CMS proton-proton ridge in ~\cite{Dumitru:2010iy}, and a quantitative description of the nearside collimated yield obtained in ~\cite{Dusling:2012iga}. 

In the first of a current series of papers~\cite{Dusling:2012cg}, the
description of long range di-hadron correlations in high multiplicity events
was significantly developed by considering both nearside and awayside
collimated contributions. In the former case, the Glasma graphs provide the
dominant contribution, while the awayside receives contributions from {\it
both} Glasma and back-to-back QCD graphs. The latter, in the high energy
kinematics of the LHC experiments, is described in the CGC EFT by BFKL dynamics~\cite{Balitsky:1978ic,Kuraev:1977fs}. We showed in this study that the BFKL dynamics, which generates gluon emissions between gluons that fragment into triggered hadrons, does well in describing the awayside spectra. The description is significantly better than PYTHIA-8~\cite{Khachatryan:2010gv}, and $2\rightarrow 4$ QCD graphs in the Quasi--Multi--Regge--Kinematics (QMRK)~\cite{Fadin:1996zv,Leonidov:1999nc}, both of which overestimate the awayside yield, especially at larger momenta. 

In the second paper in this series~\cite{Dusling:2012wy}, we applied the
Glasma+BFKL CGC framework to describe first p+Pb data on the ridge obtained at
$\sqrt{s}=5.02$ TeV/nucleon by the CMS collaboration~\cite{CMS:2012qk}. The CMS
p+Pb data had the following striking systematic features: i) a strong
dependence of the ridge yield on the number of charged particle tracks $N_{\rm track}$, with a significantly
larger signal than in p+p for the same $N_{\rm track}$, ii) a stronger $p_T$ dependence than in p+p for the same large $N_{\rm track}$, and iii) a nearside collimation for large $N_{\rm track}$ comparable to the awayside for the lower $p_T = \ptrig=\ptasc$ di-hadron windows. In ~\cite{Dusling:2012wy}, we showed that all these systematic features could be explained in the CGC framework as a consequence of a remarkable quantum intereference effect in the production of correlated gluons. 

Subsequently, both the ALICE~\cite{Abelev:2012ola} and
ATLAS~\cite{Aad:2012gla} collaborations have presented their di-hadron correlation results from the first LHC p+Pb run. The ALICE experiment
has an acceptance in $\Delta \eta$ of $|\Delta \eta|<1.8$, while the ATLAS
experiment has an acceptance of $2 < |\Delta \eta|< 5$, close to the CMS acceptance of $2<|\Delta \eta|<4$.  In addition to the LHC results, the PHENIX collaboration at RHIC have reanalyzed their deuteron-gold data at 200 GeV/nucleon and have extracted a ridge signal in very central events~\cite{Sickles-WWND}.  All three of the experiments show that when the two particle yield in peripheral collisions is subtracted from the central events, a dipole structure remains that is long range in rapidity.  This is precisely what one would anticipate in our Glasma+BFKL graph scenario, because the latter has a weak dependence on centrality, and the former has shape that is symmetric around $\Delta\phi = \pi/2$. In this work, we will address these recent analyses, and show that they can be reproduced in the CGC framework with a common set of parameters. 

This work is organized as follows. In Section 2, we will very briefly introduce our framework. This discussion will hew closely to those 
in Refs.~\cite{Dusling:2012cg,Dusling:2012wy} with many details to be found in those papers and the papers cited therein. In section 3, we will first discuss results for proton-proton collisions reported by CMS.  In addition to the ``matrix" of the collimated associated yield per trigger as a function of $\Delta\phi$ for varying windows in  $\ptrig,\;\ptasc$ for the highest multiplicity events, we will show results for different multiplicity bins as well. We will comment on some open issues here that may impact the interpretation of the p+Pb data. 
We next show a comparison of our results for p+Pb collisions for the associated yield per trigger to data from CMS, ATLAS and ALICE. 
We also present estimates for the associated yield in deuteron-gold collisions at RHIC, which is likely to be obtained soon following preliminary results presented by the PHENIX collaboration~\cite{Sickles-WWND}. In the final section, we summarize our results and discuss their implications, as well as comment on alternative interpretations of the data.  An appendix outlines the different normalization procedures followed by the different experiments, and how these can be related to our analysis and to each other. 

\section{Review of long range rapidity correlations in the CGC EFT}

\begin{figure}
\centering
\includegraphics[scale=1]{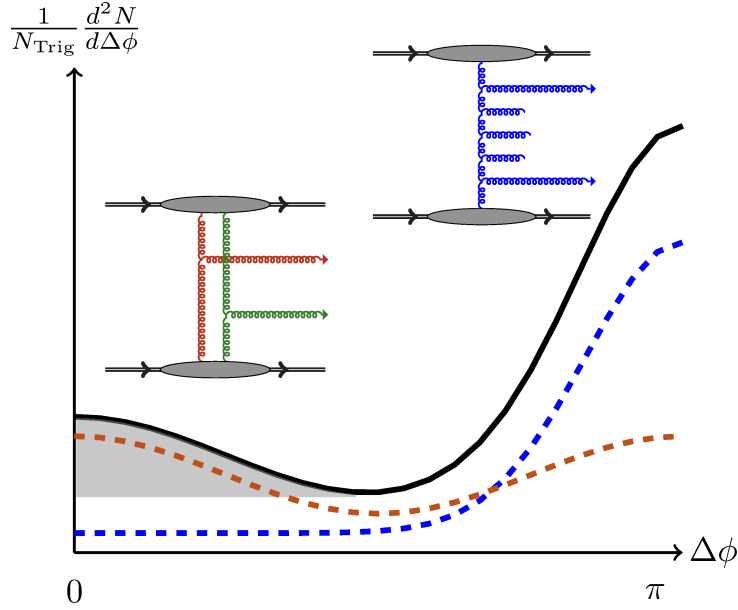}
\caption{Anatomy of di-hadron correlations.  The glasma graph on the left illustrates its 
its schematic contribution to the double inclusive cross-section (dashed orange curve). On the right is the 
back-to-back graph and the shape of its yield (dashed blue curve). The grey blobs denote emissions all the way from beam rapidities to 
those of the triggered gluons. The solid black curve represents the sum of contributions from glasma and back-to-back graphs. The shaded region represents the Associated Yield (AY) calculated using the zero-yield-at-minimum (ZYAM) procedure. Figure from ref.~\cite{Dusling:2012cg}. }
\label{fig:graph}
\end{figure}

In this section, we will review the expressions discussed in our previous papers. Except for a few clarifying details, the 
discussion is very similar to that in ~\cite{Dusling:2012cg,Dusling:2012wy}. In Fig.~(\ref{fig:graph}), as in the previous two papers, we provide a schematic sketch of the Glasma and BFKL graphs. Both of these are connected QCD graphs, and their relative power counting is given by the CGC EFT. As shown in the figure, the Glasma graphs give a collimated contribution in $\Delta \phi$ that is mirror symmetric about $\Delta \phi = \pi/2$. The ``di-jet" contribution is peaked back-to-back around $\Delta \phi = \pi$, and gives a negligible contribution on the nearside at $\Delta\phi=0$. 

The collimated contributions from all the Glasma graphs can be compactly written as~\cite{Dusling:2012wy,Dusling:2009ni}
\begin{align}
\frac{d^2N_{\rm \sl Glasma}^{\rm \sl corr.}}{d^2\pp d^2\qp dy_p dy_q}
&=  \frac{\alpha_S(\pp)\,\alpha_S(\qp)}{4\pi^{10}}\frac{N_C^2}{(N_C^2-1)^3 \,\zeta}\,\frac{S_\perp}{\pp^2\qp^2}K_{\rm glasma}\nonumber\\
&\times \left[\int_{\kp} (D_1 + D_2)+\sum_{j=\pm}\left(A_1(\pp,j\qp) + \frac{1}{2} A_2(\pp,j\qp)\right)\right] \, ,
\label{eq:Glasma-corr}
\end{align}
where we define
\begin{align}
D_1 &= \Phi_{A_1}^2(y_p,\kp)\Phi_{A_2}(y_p,\pp-\kp)
\left[\Phi_{A_{2}}(y_q,\qp+\kp)+\Phi_{A_{2}}(y_q,\qp-\kp)\right]\nonumber\,, \\
D_2 &= \Phi_{A_2}^2(y_q, \kp)\Phi_{A_1}(y_p,\pp-\kp)
\left[\Phi_{A_{1}}(y_q,\qp+\kp)+\Phi_{A_{1}}(y_q,\qp-\kp)\right]\, .
\end{align}
These four terms, called the ``single diffractive" and ``interference" graphs in \cite{Dumitru:2008wn}, constitute the leading $\pp/Q_S$ behavior.  Also included is the next order correction in $\pp/Qs$ where we have\footnote{As previously in~\cite{Dusling:2012cg}, the delta distribution is smeared as $\delta(\phi_{pq})\to
\frac{1}{\sqrt{ 2\pi \sigma}}e^{-\frac{\phi_{pq}^2}{2\sigma^2}}$, where $\Delta\phi_{p,q}=\phi_p-\phi_q$ and $\sigma=3\textrm{ GeV}/\pp$
is a $\pp$ dependent width on the order of the saturation scale. The associated yield--the integral over the near-side
signal--is insensitive to details of this smearing.}
$A_1 = \delta^2(\pp+\qp) \left[\mathcal{I}_1^2 + \mathcal{I}_2^2 + 2\mathcal{I}_3^2 \right]$, such that 
\begin{align}
\mathcal{I}_1&=\int_{\kpn{1}} \Phi_{A_1}(y_p,\kpn{1})
\Phi_{A_2}(y_q,\pp-\kpn{1}) \frac{\left(\kpn{1}\cdot
\pp-\kpn{1}^2\right)^2}{\kpn{1}^2\left(\pp-\kpn{1}\right)^2}\;,\nonumber\\
\mathcal{I}_2&=\int_{\kpn{1}}  \Phi_{A_1}(y_p,\kpn{1})
\Phi_{A_2}(y_q,\pp-\kpn{1})\frac{\left|\kpn{1}\times\pp\right|^2}{\kpn{1}^2\left(\pp-\kpn{1}\right)^2}\;,\nonumber\\
\mathcal{I}_3&=\int_{\kpn{1}}  \Phi_{A_1}(y_p,\kpn{1})
\Phi_{A_2}(y_q,\pp-\kpn{1})\frac{\left(\kpn{1}\cdot
\pp-\kpn{1}^2\right)\left|\kpn{1}\times\pp\right|}{\kpn{1}^2\left(\pp-\kpn{1}\right)^2}\;.\nonumber
\end{align}
The other contribution, $A_2$, in Eq.~(\ref{eq:Glasma-corr}) can be expressed as 
\begin{eqnarray}
A_2 =&& \int_{\kpn{1}}
\Phi_{A_1}(y_p,\kpn{1})\Phi_{A_1}(y_p,\kpn{2})
\Phi_{A_2}(y_q,\pp-\kpn{1})
\Phi_{A_2}(y_q,\qp+\kpn{1})
\nonumber\\
&\times&\frac{\left(\kpn{1}\cdot \pp-\kpn{1}^2\right)\left(\kpn{2}\cdot \pp-
\kpn{2}^2\right)+
\left(\kpn{1}\times\pp\right)\left(\kpn{2}\times\pp\right)}{\kpn{1}^2\left(\pp-
\kpn{1}\right)^2}\nonumber\\
&\times&\frac{\left(\kpn{1}\cdot \qp-\kpn{1}^2\right)\left(\kpn{2}\cdot \qp-
\kpn{2}^2\right)+
\left(\kpn{1}\times\qp\right)\left(\kpn{2}\times\qp\right)}{\kpn{2}^2\left(\qp+
\kpn{1}\right)^2}
\label{eq:double-inclusive-5}
\end{eqnarray}
where $\kpn{2}\equiv \pp-\qp-\kpn{1}$.  The above expressions are the result of including all combinatorial combinations of graphs represented by the Feynman diagram to the left in Fig.~\ref{fig:graph}. The combinatorics
is a result of different ways of averaging over strong color sources between the amplitude and complex conjugate amplitude in both projectile and target. It is important to note 
%that Eq.~(\ref{eq:Glasma-corr}) cannot be expressed as a product of an amplitude times the complex conjugate amplitude, but
that Eq.~(\ref{eq:Glasma-corr}) represents genuine quantum interference contributions, the structure of which were first outlined in~\cite{Dumitru:2008wn}. 

In equations~\ref{eq:Glasma-corr} through \ref{eq:double-inclusive-5} the only function (besides the one loop running coupling constant $\alpha_S$) is the unintegrated gluon distribution (UGD) per unit transverse area 
\begin{equation}
\Phi_A(y,k_\perp) = {\pi N_C k_\perp^2\over 2\alpha_S}\int_0^\infty dr_\perp r_\perp J_0(k_\perp r_\perp)  [1-{\cal T}_A(y,r_\perp)]^2\, 
\label{eq:unint-gluon}
\end{equation}
where ${\cal T}_A$ is the forward scattering amplitude of a quark-antiquark dipole of transverse 
size $r_\perp$ on the target $A$; it, or equivalently, the UGD, is a universal quantity that can be determined by solving the Balitsky-Kovchegov (BK) 
equation~\cite{Balitsky:1995ub,Kovchegov:1999yj} as a function of the rapidity $y=\log\left(x_0/x\right)$. Specifically, what we use for the UGDs is the rcBK equation, which includes all leading logs in x (LLx) contributions to the UGDs + running coupling next-to-leading-logs (NLLx) effects via the Balitsky prescription~\cite{Balitsky:2006wa}.  The forward scattering amplitude ${\cal T}_A(y,r_\perp)$ at the initial scale $x=x_0$ is a dimensionless function of  $r_\perp^2 Q_0^2$, where $Q_0$ is a non-perturbative scale at the initial rapidity. The saturation scale $Q_S$, defined as the transverse momentum defining the peak value of $\Phi$ on the l.h.s of eq.~(\ref{eq:unint-gluon}), is typically a larger scale even at the initial rapidity, and grows rapidly via the rcBK renormalization group equation with rapidity.  In the rcBK equation, different impact parameters in the proton/nuclear target are modeled by varying $Q_0$. The minimum-bias (median impact parameter) 
value we choose for the proton $Q_0^2 = 0.168$ GeV$^2$ (corresponding to a
$Q_S \approx 0.7~{\rm GeV}$ in the adjoint representation {\it at the initial rapidity}), is the value that gives a best fit to deeply inelastic electron-proton scattering data from HERA~\cite{Albacete:2010sy}. 

In addition to $Q_0^2$, Eq.~(\ref{eq:Glasma-corr}) has three free parameters; the first being the transverse overlap area $S_\perp$ is fixed separately for p+p, d+Au and p+Pb collisions and will be discussed in detail later.  Once $S_\perp$ is fixed \footnote{As shown later, this parameter drops out in the per trigger yield, but is present in $\Ntrk$.} for a given system, the centrality dependence is controlled entirely though the choice of initial saturation scales for the projectile and target.  A second parameter, held fixed to the same value in both p+p and p+A collisions, is the non-perturbative constant $\zeta=1/6$ specifying the correction to the $k_T$ factorized UGD description due to soft multigluon interactions. It is independently constrained by fits to empirical p+p multiplicity distributions within the $k_\perp$ factorization approximation for multi-gluon production~\cite{Tribedy:2010ab,Tribedy:2011aa} and by real time classical Yang-Mills computations~\cite{Lappi:2009xa,Schenke:2012hg}. We note further that the collimated structures seen in the perturbative {\it classical} computations persist in the full non-perturbative classical results, thereby lending confidence that the latter primarily renormalize the amplitude of the former. The third parameter $K_{\rm Glasma}$ we will discuss shortly. 

The framework of Glasma graphs is based on the factorization theorems for
``dense-dense" systems\footnote{This corresponds to the situation where both
projectile and target are highly occupied. This assumption is very plausible in
the high energy kinematics of the LHC, because i) high multiplicity events
trigger on very central impact parameters in both projectile and target, and
ii) because both gluons are produced at large rapidities relative to the beam
rapidity, allowing significant room for small $x$ evolution.}  derived
in~\cite{Gelis:2008sz}, which include leading log corrections to all orders in
perturbation theory (so called LLx approximation) as well as all leading
multiple scattering contributions~\footnote{For a recent derivation of correlated two gluon production in {\em dilute-dense} systems, where one of the projectiles is not saturated while the other is, see ~\cite{Kovchegov:2012nd}.}.  As the full expression is very cumbersome, a Gaussian truncation is employed in~\cite{Dusling:2009ni}, where Eq.~(\ref{eq:Glasma-corr}) was first derived.  The Gaussian truncation was shown in \cite{Dumitru:2011vk} to be a very good approximation to the full JIMWLK evolution.  In addition, it is assumed that $\Qs < \kp$, in order to obtain the expression in terms of UGDs (unintegrated gluon distributions).  We emphasize that the resulting expression cannot be interpreted simply as the product of UGDs with matrix elements, but combines LLx contributions to each.  

We now consider the double inclusive distribution from the back-to-back BFKL 
graphs \footnote{In the power counting of the CGC EFT, this contribution would only appear at NLLx in the dense-dense 
saturation limit. However, because the collimated LLx Glasma graph contributions are $N_c^2$ suppressed, the BFKL back-to-back contributions are competitive on the awayside. Also, as one goes away from the dense limit, the power couting shifts rapidly and the back-to-back graphs become huge relative to the glasma graphs.} shown in Fig.~\ref{fig:graph}. The double inclusive multiplicity can be expressed as~\cite{Colferai:2010wu,Fadin:1996zv}
\begin{eqnarray}
\label{eq:BFKL}
\frac{d^2N_{\rm \sl BFKL}^{\rm \sl corr.}}{d^2\pp d^2\qp dy_p dy_q} &=& \frac{32\,\nc\,
\alpha_s(\pp)\,\alpha_s(\qp)}{ (2\pi)^8 \,\cf}\,\frac{\sp}{\pp^2\qp^2}K_{\rm bfkl}\\
&\times&\int_{\kpn{0}} \int_{\kpn{3}}
\Phi_A(x_1,\kpn{0})\Phi_B(x_2,\kpn{3})\,\mathcal{G}(\kpn{0}-\pp,\kpn{3}+\qp,y_p-y_q)\nonumber
\end{eqnarray}
where $\mathcal{G}$ is the BFKL Green's function
\begin{eqnarray}
\mathcal{G}(\qpn{a},\qpn{b},\Delta y)=\frac{1}{(2\pi)^2}\frac{1}{(\qpn{a}^2 \qpn{b}^2)^{1/2}}\sum_n e^{in\overline{\phi}}\int_{-\infty}^{+\infty} d\nu\textrm{ } e^{\omega(\nu,n)\Delta y}e^{i\nu\ln\left(\qpn{a}^2/\qpn{b}^2\right)}\textrm{   } \, .
\label{eq:BFKL-Green}
\end{eqnarray}
Here $C_F = (\nc^2-1)/2 \nc$, ${\omega(\nu,n)=-2\overline{\alpha}_s\,
\textrm{Re}\left[\Psi\left(\frac{|n|+1}{2}+i\nu\right)-\Psi(1)\right]}$ is the
BFKL eigenvalue, where  $\Psi(z)= d\ln\Gamma(z)/dz$ is the logarithmic
derivative of the Gamma function. Further, we have $\overline{\alpha}_s\equiv
\nc\,\as\left(\sqrt{\qpn{a}\qpn{b}}\right)/\pi$ and 
$\overline{\phi}\equiv \arccos\left(\frac{\qpn{a}\cdot \qpn{b}}{\vert\qpn{a}\vert\textrm{ }\vert \qpn{b}\vert}\right)$. 

In the description of the away-side jet in the above BFKL framework, the UGD evolution, as for the Glasma graphs is described by the rcBK equation. NLLx corrections to the back-to-back graphs have been computed in \cite{Colferai:2010wu}.  It was demonstrated there that the NLLx correction to the $\Delta\phi$ independent pedestal is a large one (a factor 2 to 3).  However, the NLLx contribution to the collimated $\langle\cos( \Delta \phi)\rangle$ and $\langle\cos(2\Delta\phi)\rangle$ moments (as also confirmed in ~\cite{Caporale:2012qd}), which are the quantities of interest here, is 10-30\%; further, we expect our inclusion of running coupling in Eq.~(\ref{eq:BFKL-Green}) will potentially account for a good fraction of this correction. Based on the results in these works, it is reasonable to conclude that the BFKL contribution to the collimated yield has 10-30\% uncertainties.

\begin{figure}
\includegraphics[width=6in]{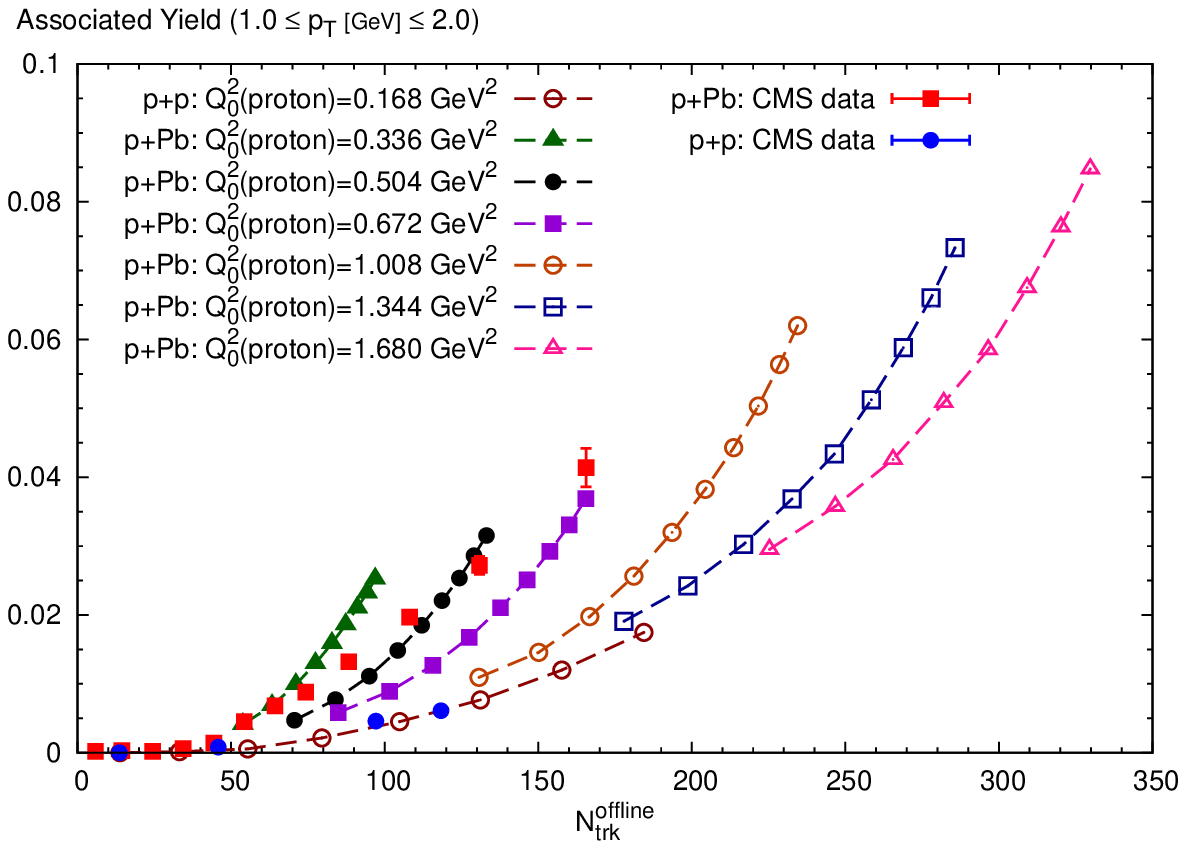}
\caption{The nearside yield per trigger as a function of $\Ntrk$ (specific to
the acceptance of the CMS experiment) for $1\leq p_T \leq 2$, for $p_T=\ptrig=\ptasc$.  The open circles (brown) are the computed Glasma graph yield for increasingly rare proton initial saturation scales in multiples $\Npp$ of $Q_0^2 = 0.168$ GeV$^2$. The filled circles (blue) are proton-proton data at $\sqrt{s}=7$ TeV from the CMS collaboration. Each of the p+Pb curves corresponds, for fixed $Q_0^2 ({\rm proton})$, to the increasing yield with a larger number of participants in the nucleus (in increments of two, from $\Npb=6$ to $\Npb=22$), where $\Npb$ denotes the initial saturation scale in the Pb nucleus through the relation $Q_0^2{\rm (lead)} =
\Npb\cdot 0.168~\rm{GeV}^2$.  The three curves with open symbols are new computations for higher multiplicities.  In this case the number of participants in the nucleus increases in increments of two up to $\Npb=26$.  The red squares are CMS data for proton-lead collisions at $\sqrt{s}=5.02$ GeV/nucleon.  Similar curves are obtained for the $\Ntrk$ appropriate in the acceptance of the other experiments.
}
\label{fig:multi_revised}
\end{figure}

As shown in Fig.~(\ref{fig:graph}), Eq.~(\ref{eq:BFKL}) gives a collimated $\Delta \phi$ contribution exclusively on the away side, peaked at 
$\Delta\phi = \pi$, while Eq.~(\ref{eq:Glasma-corr}) gives a ``dipole" $\cos(2\Delta\phi)$-like contributions with maxima at 
$0$ and $\pi$.  It should be noted that the behavior of both sets of graphs for any given window in $p_T$ and rapidity is strongly influenced  by the evolution of the UGDs which are common to both and provide a non-trivial constraint on the relative contributions of each to the description of data. It's the interplay between these contributions with varying $Q_0$ in projectile and target that describes the systematics of the proton-proton and proton-lead data, that we shall discuss quantitatively shortly.

The  K-factors used
throughout this work, $K_{\rm Glasma}$ and $K_{\rm BFKL}$, are introduced to take into account not only uncertainties
in higher order computations but also acceptance corrections and uncertainties
in the choice of fragmentation functions.
For simplicity, we
will take them to be equal to each other in each process, but will use
different values, as stated at the appropriate juncture, for p+p and p+A.
There is no reason a priori why all these should be the same in p+p and p+A.
Also, modulo a better understanding of the multiplicity distribution in p+A,
some of the uncertainties in K factors could be absorbed in the initial
saturation scale $Q_0$ or vice versa, corresponding to slightly different
number of participants. In addition to data on multiplicity distributions in
the rapidity window of interest,  uncertainties on fragmentation functions in
particular can be constrained by forthcoming data from the LHC on single
particle spectra at forward rapidities.

\begin{figure}[htb]
\includegraphics[width=1.0\textwidth]{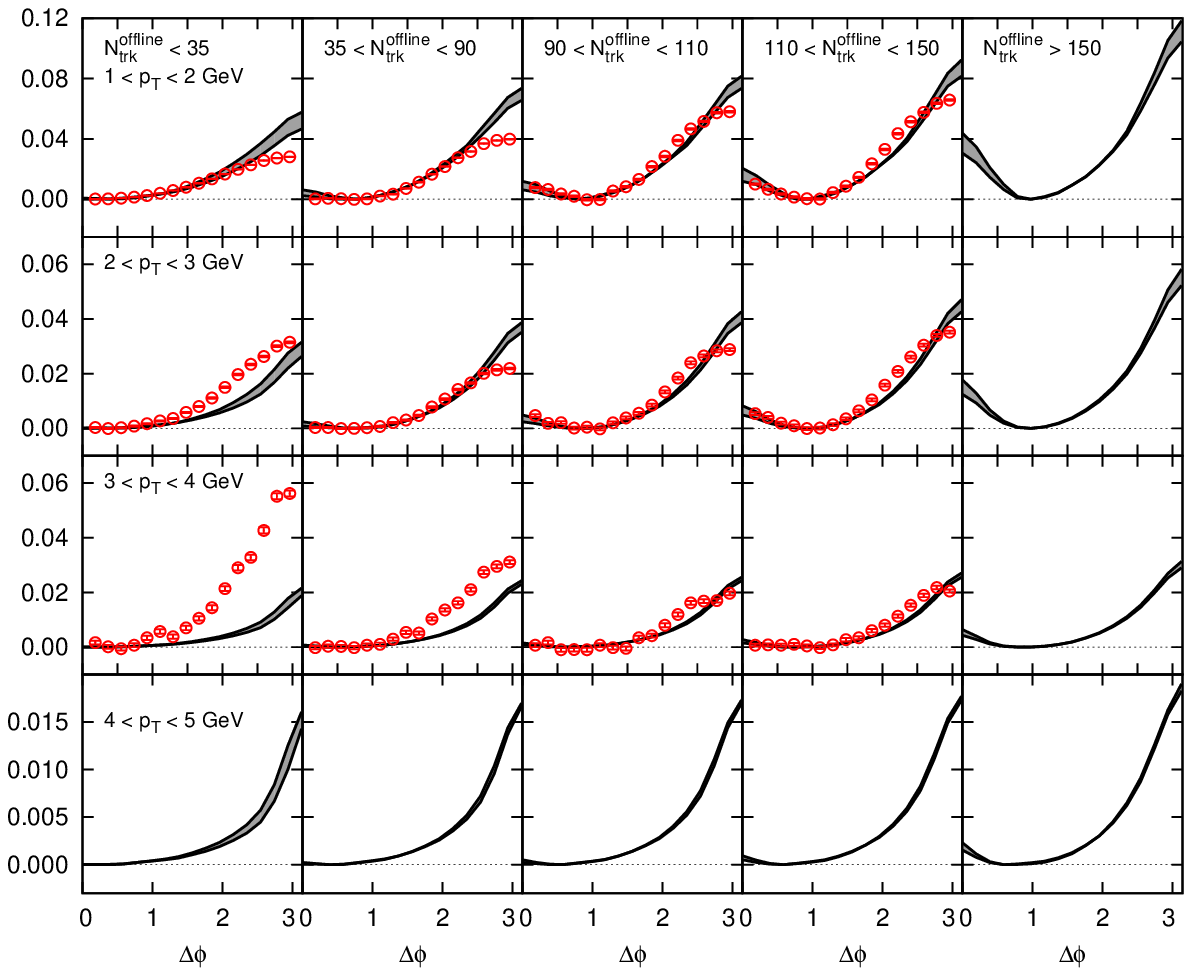}
\caption{Long-range $\left(2\leq \vert\Delta\eta\vert \leq 4\right)$
per-trigger-yields $\left(1/\Ntrig d^2N/d\Delta\phi\right)$ of charged hadrons
as a function of $\vert \Delta\phi\vert$, for p+p collisions at
$\sqrt{s}=7$~TeV.  Data are from the CMS collaboration. The lower (upper) curves
indistinguishable in some windows, correspond to the following: i) $\Ntrk<35$: $\Npp=1,2$ ii) $35<\Ntrk<90$: $\Npp=3,4$ iii) $90<\Ntrk<110$: $\Npp=4,5$ iv) $110<\Ntrk<150$: $\Npp=5,6$ v) $\Ntrk>150$: $\Npp =7,8$. 
}
\label{fig:CMS_pp}
\end{figure}

\begin{figure}[]
\includegraphics[width=1.0\textwidth]{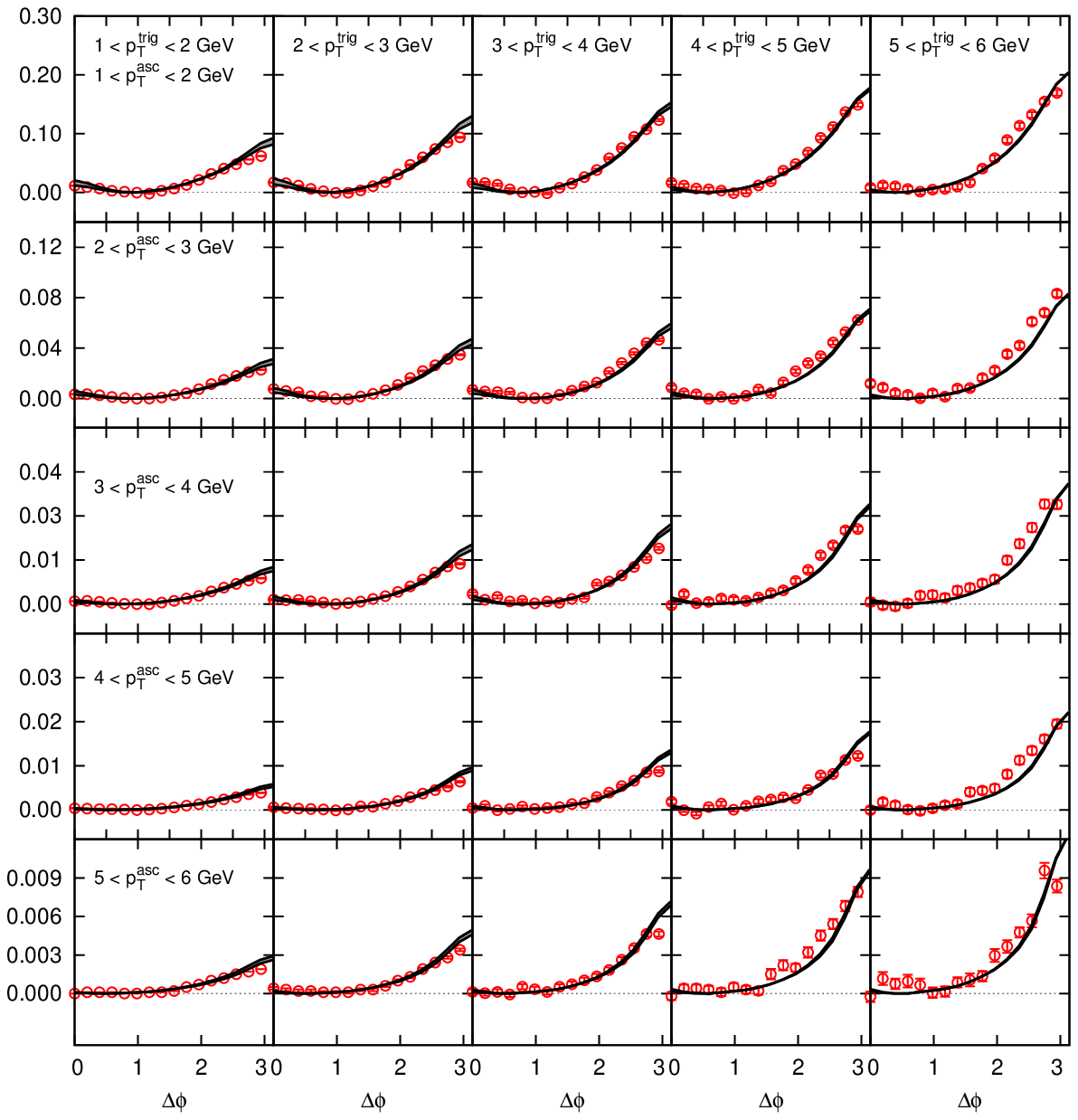}
\caption{Long-range $\left(2\leq \vert\Delta\eta\vert \leq 4\right)$ per-trigger-yields $\left(1/\Ntrig d^2N/d\Delta\phi\right)$ of charged hadrons as a function of $\vert \Delta\phi\vert$, from high multiplicity $\left(\Ntrk \geq 110\right)$ pp collisions at $\sqrt{s}=7$~TeV.  Data are from the CMS collaboration. The lower (upper) curves, indistinguishable in some windows, correspond to 
$\Npp =5,6$.
}
\label{fig:CMS_pp2}
\end{figure}

\section{Quantitative study of di-hadron correlations at wide rapidity separations}

For the analysis of the LHC proton-proton and proton-lead data, as well as the RHIC deuteron-gold data, we must be able to make a reasonable estimate of the centrality class based on the total charge particle multiplicity. In the CGC framework, the single inclusive gluon distribution is defined as~\cite{Dusling:2012iga}
\begin{align}
\frac{\ud N_1}{\ud y_p\ud^2\pp }
=\frac{\alpha_s N_C}{4\pi^6 (N_C^2-1)}\frac{S_\perp}{\pp^2}
\int_{\kp}\!\!\!
\Phi_{A_1}(y_p,\kp)\,\Phi_{A_2}(y_p,\pp-\kp)\,,
\label{eq:single-incl}
\end{align}
where the $\Phi$'s are the UGDs defined previously. The number of charged hadron tracks is then defined as 
\begin{align}
\Ntrk=\kappa_g \int_{-y_{\rm accept}-y_{\rm shift}}^{+y_{\rm accept}-y_{\rm shift}}\! \!\!\!\! d\eta \!\!\int_{\ptmin}\!\!\!\! d^2\pp \frac{dN}{d\eta \,d^2 \pp}\left(\pp\right)\, .
\label{eq:ntrk}
\end{align}
where $y_{\rm shift}=0.465$ is the shift in rapidity in the center-of-mass frame in asymmetrical p+Pb collisions towards the lead fragmentation region, $y_{\rm accept}$ is the maximal laboratory rapidity of a given detector and $p_{T,\rm min}$ is the minimal transverse momentum of measured charged tracks in the detector \footnote{The number of charged tracks $N_{\rm trk}$ is 
distinguished from $\Ntrk$ because the former includes an extrapolation down to $p_{T,\rm min}=0$.} . Since Eq.~(\ref{eq:single-incl}) corresponds to the single inclusive gluon multiplicity, and since the bulk of particle production is soft, we have to introduce a 
gluon liberation factor $\kappa_g$, which specifies, assuming parton-hadron duality, the conversion of gluons to charged hadrons.  In Eq.~(\ref{eq:ntrk}), the combination of the transverse overlap area $S_\perp$ times $\kappa_g$ is separately fixed for p+p and p+Pb to give the best description of the data.  For minimum bias proton--proton collisions ($Q_0^2=0.168$ GeV$^2$ in both protons) this corresponds to $\Ntrk=14$.

This value, ($\kappa_g S_\perp)$, is subsequently held fixed to determine $\Ntrk$ as $Q_0^2$ (the saturation scale at the initial rapidity for $x$ evolution) in both the proton and lead nucleus is varied.  Admittedly, the constant $(\kappa_g S_\perp)$ provides a lot of freedom in the centrality selection, but it can only be constrained as further data on single inclusive quantities in the rapidity ranges of interest become  available.  The uncertainty in the value of $(\kappa_g S_\perp)$ results in a rescaling of the x-axis of Fig.~\ref{fig:multi_revised}; Neither $\kappa_g$ or $S_\perp$ enter into the computation of the associated yield.  As discussed later, the interaction cross section is on the order of the size of the proton and not the nucleus. 
%may vary with both beam energy and centrality.  We expect the uncertainty due to these corrections to be small but not negligible.  As shown in \cite{Tribedy:2010ab} the value of $S_\perp$ varies only logarithmically with beam energy.  Furthermore, 

A part of the analysis of the di-hadron data requires the calculation of the number of trigger particles, defined here as
\begin{align}
\Ntrig=\int_{-y_{\rm accept}-y_{\rm shift}}^{+y_{\rm accept}-y_{\rm shift}}\! \!\!\!\! d\eta \!\!\int_{p_T^{\rm min}}^{p_T^{\rm max}}\!\!\!\! d^2\pp\!\!\int_{z_0}^1 \!\!  dz \frac{D(z)}{z^2} \frac{dN}{d\eta \,d^2 \pp}\left(\frac{p_{\textrm{T}}}{z}\right)\, ,
\label{eq:ntrig}
\end{align}
where $p_{T,\rm min}$, $p_{T,\rm max.}$ denote the width of the $p_T$ window wherein triggered particles are selected. Since the triggered hadrons are semi-hard, one takes account of the possibility that they were generated by the fragmentation of higher $p_T$ gluons by fragmentation functions. These are chosen, as in \cite{Dusling:2012cg}, to be the NLO KPP  parametrization~\cite{Kniehl:2000fe} of the fragmentation function of gluons to charged hadrons. 

The double inclusive multiplicity of charged hadrons is computed as
\begin{align}
\label{eq:dihadron}
&\frac{d^2N}{d\Delta \phi} = \int_{-y_{\rm accept}-y_{\rm shift}}^{y_{\rm accept}-y_{\rm shift}} \!d\eta_p  \,d\eta_q \,\, {\cal A}\left(\eta_p,\eta_q\right) \\
&\!\!\times\int_{p_T^{\rm min}}^{p_T^{\rm max}} \frac{dp_T^2}{2} \int_{q_T^{\rm min}}^{q_T^{\rm max}}\frac{ d q_T^2}{2}\;\int d\phi_p \int d\phi_q\; \delta\left(\phi_p-\phi_q-\Delta\phi\right) \nonumber\\
&\!\!\times \int_{z_0}^1\!\! dz_1 dz_2 \frac{D(z_1)}{z_1^2}\, \frac{D(z_2)}{z_2^2}
 \frac{d^2N_{}^{\rm \sl corr.}}{d^2\pp d^2\qp d\eta_p d\eta_q}\left(\frac{p_{\textrm{T}}}{z_1},\frac{q_{\textrm{T}}}{z_2},\Delta\phi \right)\nonumber
\end{align}
Bounds on the range of the trigger and associated hadron momenta are denoted respectively as $p_T^{\rm min (max)}$ and $q_T^{\rm min (max)}$. Likewise, $\Delta\eta_{\rm min}(\Delta \eta_{max})$ denote the pseudo-rapidity separation
between the measured hadrons for a given detector\footnote{Replacing the
rapidity $y$ with the pseudo-rapidity $\eta$  is a good approximation for the
$p_T$, $q_T$ of interest.}. The acceptance\newline ${\cal
A}\left(\eta_p,\eta_q\right)$  
%\equiv \theta\left( \vert\eta_p -\eta_q\vert -\Delta\eta_{min}\right)\, \theta\left(\Delta\eta_{\max} - \vert
%\eta_p-\eta_q\vert\right)/{\cal B}$ where\newline ${\cal B}_{\rm CMS}=2\int_{\Delta\eta_{\rm min}}^{\Delta \eta_{max}} d\Delta\eta \left(1-\Delta\eta/{2 \eta_{\rm accept}}\right)$ 
takes into account the acceptance of the uncorrelated background. The different treatments of this acceptance function between the ATLAS, CMS and ALICE experiments are discussed in the Appendix. The PHENIX analysis is identical to that of ATLAS. 

The collimated associated yield is computed using the Zero-Yield-at-Minimum (ZYAM) procedure,
\begin{equation}
\label{eq:zyam}
\textrm{Assoc. Yield} = \frac{1}{\Ntrig}\int_0^{\Delta\phi_{\rm min.}} \!\!\!\!
d\Delta\phi\left(\frac{d^2N}{d\Delta\phi}-\left.\frac{d^2N}{d\Delta\phi}\right|_{\Delta\phi_{\rm
min}}\right)
\end{equation}
where $\Delta\phi_{\rm min.}$ is the angle at which the two particle correlation strength is minimal. An important point to note is that the transverse overlap area $S_\perp$ cancels out between the numerator and denominator in the r.h.s eliminating a source of uncertainty in di-hadron spectra.  

After these preliminaries, we are now ready to discuss our results. In Fig.~(\ref{fig:multi_revised}), we plot the integrated associated nearside yield per trigger (obtained from Eqs.~(\ref{eq:dihadron}) and (\ref{eq:zyam})) versus $\Ntrk$ as determined in Eq.~(\ref{eq:ntrk})  for $1\leq p_T \leq 2$, for $p_T=\ptrig=\ptasc$. The associated yield in p+p collisions in the computation is shown by the open brown circles that are connected by dashed brown lines. These open circles correspond to results in integer multiples of $Q_0^2 = 0.168$ GeV$^2$, which as noted previously is the saturation scale at the initial $x_0$ in fits of the rcBK equation to inclusive deeply inelastic scattering data at HERA. The blue filled circles correspond to the CMS proton-proton data for the nearside associated yield in the different $\Ntrk$ windows specified by the collaboration. We observe that they lie nicely on this curve. This then helps us identify the range in $Q_0^2$ (proton) that matches the $\Ntrk$ centrality selection in the experiment to be discussed shortly in comparisons to the detailed matrix of the collimated yield versus $\Delta \phi$. 

The sole inputs for the trajectories shown in Fig.~(\ref{fig:multi_revised}) are $Q_0^2$(proton)=$\Npp \cdot 0.168$ GeV$^2$ and $Q_0^2$(lead) =$\Npb \cdot 0.168$ GeV$^2$. Each of the curves corresponds to a
fixed $Q_0^2$ in the proton of $0.168 - 0.672~{\rm GeV}^2$ (or $\Npp$=1-4) representing estimates of these quantities from median (``min.~bias") impact parameters in the proton to the very central impact parameters respectively that are triggered in high multiplicity events. The trajectories corresponding to each of these proton $Q_0^2$ show how the yield increases with a larger number of participants in the nucleus.  Because the CMS p+Pb data have the same $\Delta \eta$ and centrality selections, we plot these as well. The message one draws from interpreting this figure is that one is not only accessing rarer Fock configurations (at a given impact parameter) in the nucleus with increasing $\Npb$ but also rare Fock states in the proton represented by the increasing 
$\Npp$.  

Fig.~(\ref{fig:multi_revised}) captures the essence of variations in the Glasma
yield with $\Ntrk$ and $\Npb$. It shows clearly that the yield in central p/d+A
collisions is significantly enhanced relative to the yield in p+p collisions for
the same $\Ntrk$. The underlying physics behind these curves is the quantum
interference of the UGDs and the sensitivity to the spectrum of gluons in the
projectiles. A detailed analysis of these systematics was performed in~\cite{Dusling:2012wy} and we refer the interested reader to the discussion there \footnote{A curve similar to Fig.~(\ref{fig:multi_revised}) was shown in ~\cite{Dusling:2012wy} without the proton-proton and proton-lead data points. As we will discuss below, the reanalysis of data from the various experiments leads to slight shifts in the points on this plot relative to those previously.}.

We will now move forthwith to a comparison to data on the collimated nearside and awayside yields in proton-proton and proton-nucleus collisions at the LHC. In \cite{Dusling:2012cg}, we presented a comparison to the high multiplicity CMS data for 7 TeV proton-proton collisions. Subsequently, the CMS collaboration presented a detailed matrix in several centrality windows of the associated yield as a function of $\Delta \phi$ \cite{CMS:2012qk}. We also realized that the normalization of the CMS acceptance was different from what we assumed it to be--see the appendix for a detailed discussion. 

With this additional information, we have performed  here a reanalysis of the CMS proton-proton data. The results are shown in Figs.~(\ref{fig:CMS_pp}) and~(\ref{fig:CMS_pp2}) and correspond to $p_T=\ptrig=\ptasc$. The comparison to data in Fig.~(\ref{fig:CMS_pp}) is shown for the first time while a comparison to~(\ref{fig:CMS_pp2}) was shown previously. The key difference to our previous comparison is that our fits for p+p are performed with a common $K$-factor for both Glasma and BFKL graphs $K_{\rm Glasma}=K_{\rm BFKL} =1.5$. The bands in the plot correspond to different choices in $Q_0^2$ for the different track selections which are constrained to reproduce the $\Ntrk$ in these windows. What these are can be deduced from the x-axis of  Fig.~(\ref{fig:multi_revised}) as discussed previously. We see that the agreement of the theory curves to the data in the higher multiplicity windows is quite good. 

However, there is a significant discrepancy in the two lowest multiplicity windows, especially at higher $p_T$. The reasons are two fold. Firstly, our formalism for both Glasma and back-to-back contributions is less valid for these multiplicities, where more peripheral impact parameters in the proton are accessed. However, a more important reason may be that the experiments require 
at least two particles in each bin to extract a di-hadron signal. Our theory computation (in absence of Monte-Carlo simulations that 
are challenging for interference graphs) imposes no such restriction. In lower multiplicity windows, and at higher $p_T$, this is likely to provide a significant correction, just as seen in the theory comparison to data. 

Results for the collimated yield versus $\Delta \phi$ matrix in $\ptrig,\ptasc$ for the high multiplicity $\Ntrk \geq 110$ window are shown in Fig.~(\ref{fig:CMS_pp2}). The difference to the plot shown previously in ~\cite{Dusling:2012cg} is the common $K$ factor for both BFKL and Glasma graphs of $K=1.5$. Further, the $Q_0$'s corresponding to the high multiplicity window are higher. As noted earlier, this change was driven by a better understanding of the normalization of the different experiments, in this case CMS, subsequent to our previous papers. The $Q_0$'s chosen are seen in Fig.~(\ref{fig:multi_revised}) to better represent the centrality classes in multiplicity for the detailed matrix comparison. Given these noted changes, the agreement with the data over nearly 400 data points is remarkably good.  While the Glasma signal is small in the data, the back-to-back correlation is significant, and it is striking that the 
BFKL graph captures its systematics so well \footnote{Note in particular the different y axes as one goes down the matrix, denoting very different amplitudes in the different windows.}. As we argued in \cite{Dusling:2012cg}, the multi-regge (QMRK) $2\rightarrow 4$ di-jet contribution without BFKL evolution between triggered gluons shows a significantly larger collimation, a conclusion that remains unchanged. The data in our view is demonstrating decorrelation of the back-to-back di-hadron signal due to QCD evolution 
{\it a la BFKL} between the triggered hadrons. 

We now turn to a discussion of a comparison of our Glasma+BFKL framework to first data from proton-lead collisions at the LHC. In ~\cite{Dusling:2012wy}, we made a comparison to first data from the CMS collaboration~\cite{CMS:2012qk}. In this paper, we will revisit that comparison for the reasons articulated above. Further, we will make quantitative comparisons with data from ALICE and 
ATLAS within their distinct experimental acceptance. As noted, a comparison of these is discussed in the appendix. Though first PHENIX data on very central deuteron-gold data has been presented at a conference~\cite{Sickles-WWND}, we are unable to make a direct comparison because the quantity presented, unlike the collimated yield, is sensitive to the combinatorial background. We do however make a prediction for the collimated yield. 

To simulate the p+Pb collision, we  vary $Q_0^2$ at the initial rapidity scale in the proton and lead nuclei. All other parameters are
 the same, with the exception of $K_{\rm Glasma}=K_{\rm BFKL} =1$, as opposed to $1.5$ for proton-proton collisions. There is no reason a priori that $K$ factors should be the same in the p+p and p+Pb case. It is conceivable that the dense-dense factorization used here is more applicable in the latter with smaller higher order corrections. 
In our treatment, the proton $Q_0^2$ is varied in multiples $\Npp$ of the ``minimum bias" value of $Q_0^2=0.168$ GeV$^2$ to simulate proton-lead collisions that select more central impact parameters in the proton, where the gluon density is considerably higher than the gluon density for the median impact parameter corresponding to minimum-bias events. On the lead side, as noted previously, the initial saturation scale in lead is $Q_0^2= \Npb \cdot 0.168~{\rm GeV}^2$, where $\Npb$ denotes the number of  participants (color charge probed) in the lead nucleus. 

\begin{figure}
\includegraphics[width=5in]{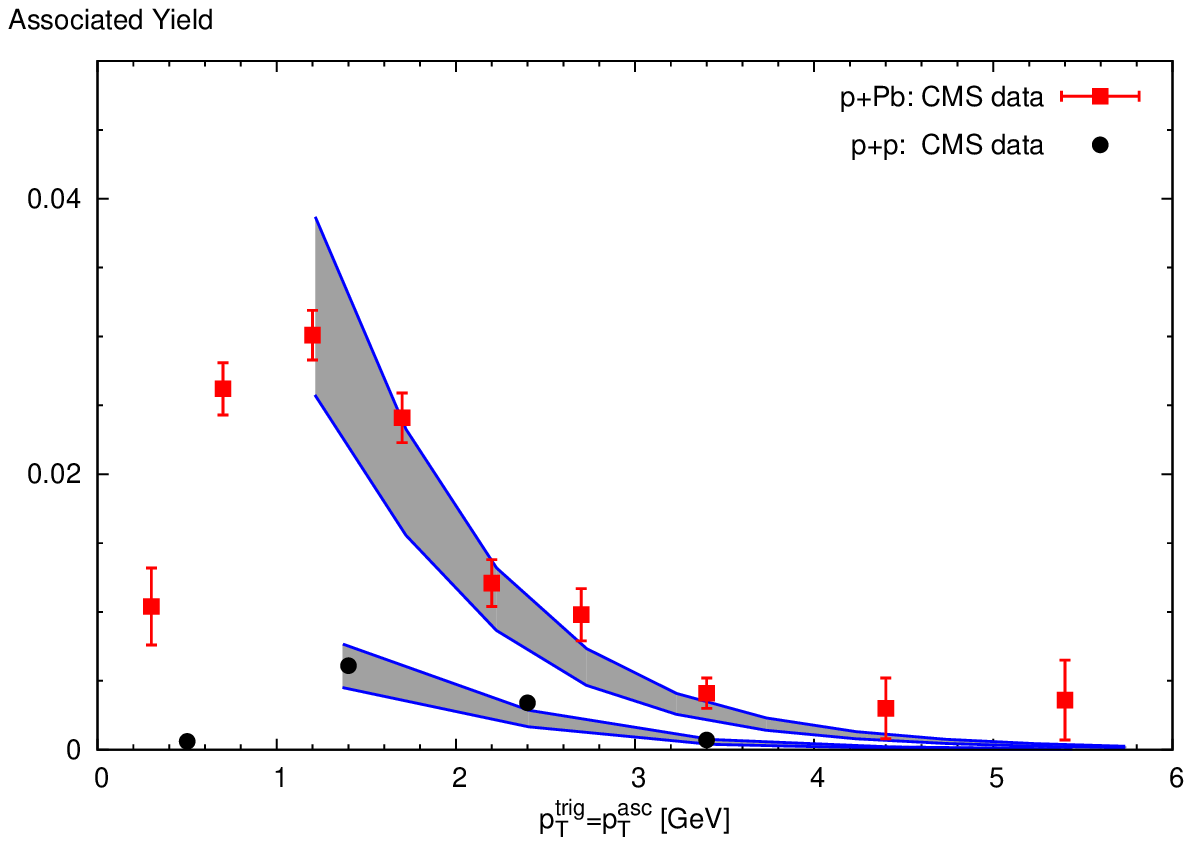}
\caption{The $p_T$ ($\ptrig=\ptasc$) dependence of the associated yield in proton-lead and proton-proton collisions.  The data here are for $\Ntrk \geq 110$. With the guidance from Fig.~(\ref{fig:multi_revised}), the  proton-lead centrality band correspond to ($\Npp$,$\Npb$) of (3,22) (upper curve) and (4,14) (lower curve). The proton-proton curves correspond to $\Npp$ = 5 and 6. See text for further explanation.} 
\label{fig:ay_pPb}
\end{figure}

Specifically, to compare to the nearside collimated yield data in
Fig.~(\ref{fig:ay_pPb}), the lower and upper curves correspond to 
$Q_0^2$ values that are respectively five to six times the minimum bias value. As seen from the x-axis of Fig.~(\ref{fig:multi_revised}), they provide a reasonable estimate of the $\Ntrk$'s one can estimate contributing to the centrality cut 
$\Ntrk \geq 110$. By the same logic, for the same centrality cut in proton-lead collisions, the guidance afforded by Fig.~(\ref{fig:multi_revised}) suggests ($\Npp$,$\Npb$) =(3,22) (upper curve) and (4,14) (lower curve). These are of course estimates, but we have checked that small variations of these do not significantly widen the uncertainty band. 

We see from Fig.~(\ref{fig:ay_pPb}) that the nearside yield from the Glasma graphs, within theoretical uncertainties, is able to account for the $p_T=\ptrig=\ptasc$ dependence of the CMS measurement for the high multiplicity window in both proton-proton and proton-lead collisions. In particular, it naturally explains the factor of six enhancement in proton-lead to proton-proton collisions in the $p_T=1-2$ GeV window where the yields are the largest. While some final state rescattering cannot be ruled out, we believe it will be difficult to account for this large factor in hydrodynamic models~\cite{Bozek:2012gr}. In our picture, the signal is due to the 
quantum interference of the unintegrated gluon distributions in the projectile and target. As shown in \cite{Dusling:2012wy}, the contribution of this overlap scales approximately as $\Ntrk\cdot \Npb$, as a result of which one obtains the trajectories shown in Fig.~(\ref{fig:multi_revised}). 

\begin{figure}[htb]
\includegraphics[width=1.0\textwidth]{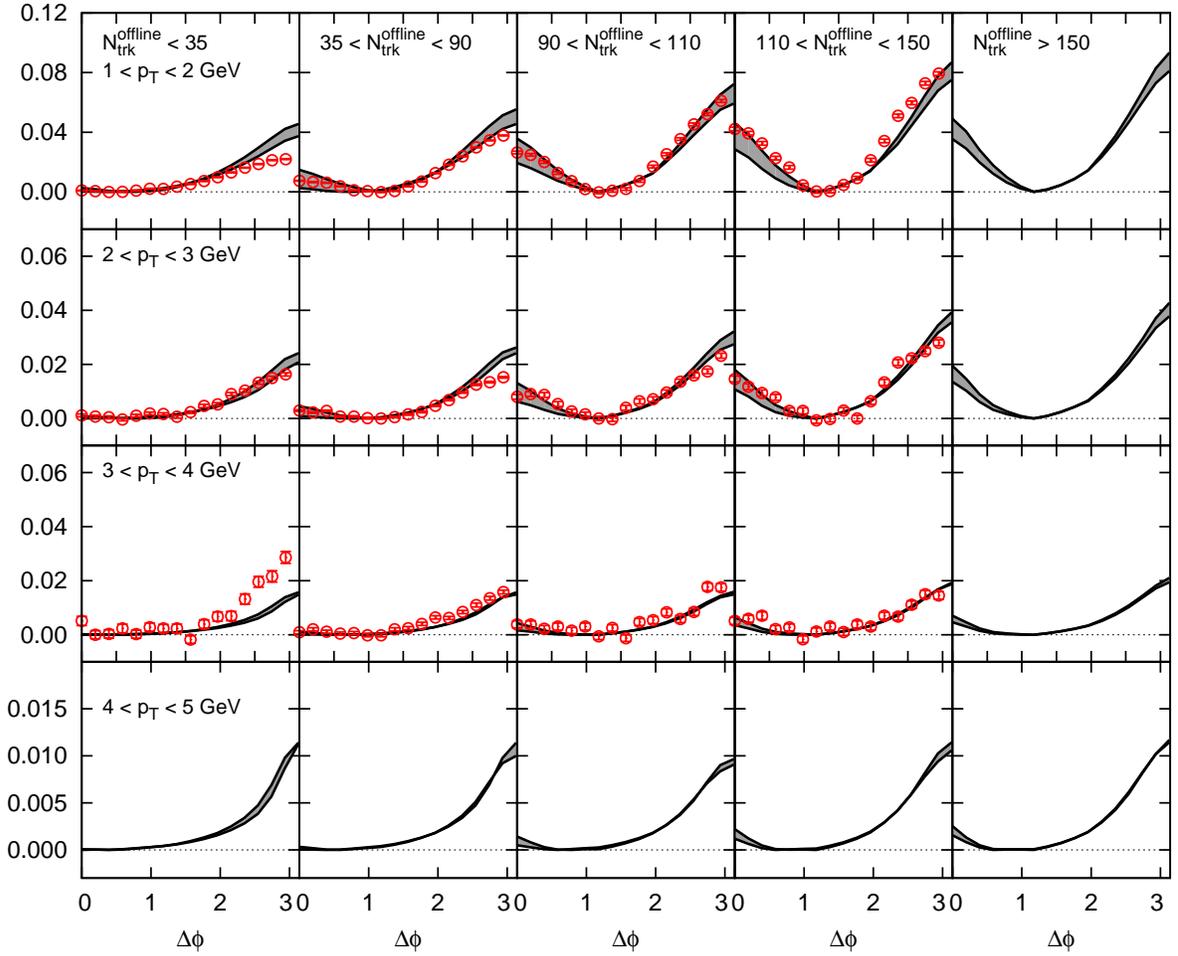}
\caption{Long-range $\left(2\leq \vert\Delta\eta\vert \leq 4\right)$ per-trigger-yields $\left(1/\Ntrig d^2N/ d\Delta\phi\right)$ of charged hadrons as a function of $\vert \Delta\phi\vert$ in different $p_T$ and multiplicity bins for pPb collisions at $\sqrt{s}=5.02$~TeV with the CMS experiment.  The $p_T$ selection corresponds to both particles in the pair.  The centrality dependence of the theory curves are controlled by the choice of initial saturation scale in the target and projectile. These are  i) $\Ntrk<35$: ($\Npp,\Npb$) = (1, 3) (lower curve),  (2, 6) (upper curve), ii) $35<\Ntrk<90$: (2, 6) (lower), (2, 12) (upper), iii) $90<\Ntrk<110$: (2,14) (lower),  (2,22) (upper), iv) $110<\Ntrk<150$: (3, 22) (lower), (4, 14) (upper), and v) $\Ntrk>150$: (4, 16) (lower), (4,20) (upper).
}
\label{fig:matrix_CMS_pPb}
\end{figure}

A more detailed comparison of the CGC EFT framework to data is obtained in Fig.~(\ref{fig:matrix_CMS_pPb}). Here the CMS proton-lead data for the collimated yield as a function of $\Delta \phi$ are shown for a number of windows in $\Ntrk$ and in $p_T=\ptrig=\ptasc$. The saturation scales for each $\Ntrk$ centrality window are estimated with the guidance from Fig.~(\ref{fig:multi_revised}). As in the proton-proton case, we see that the agreement is quite good, especially in the higher multiplicity windows. Again, as in the p+p case, we see the most significant deviation from data is the underprediction in the lower multiplicity bins at higher $p_T$. We believe this to have the same underlying cause as in the proton-proton case; a trigger-bias in the experiment where only events containing at least two hadrons in the $p_T$ windows of interest are included in the averaging.  At lower $p_T$ and higher multiplicities, where the yield of charged particles is larger, this effect becomes insignificant.

%%%%%%%%%%%%%%%%%%%%%%%%%%%%%%%%%%%%%%%%%%%%%%%%%%%%%%%%%%%%%%%%%%%%%%%%%
\begin{figure}
\includegraphics[width=2.4in]{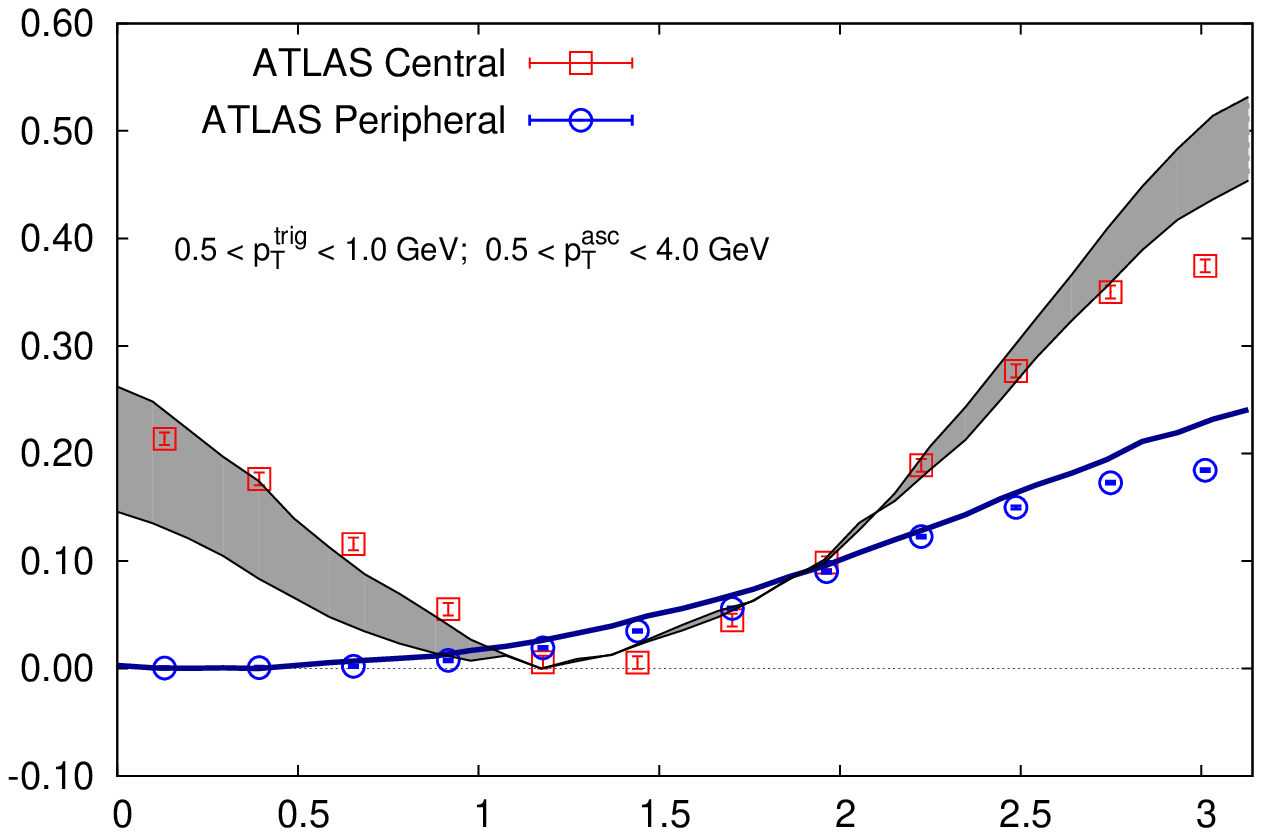}
\includegraphics[width=2.4in]{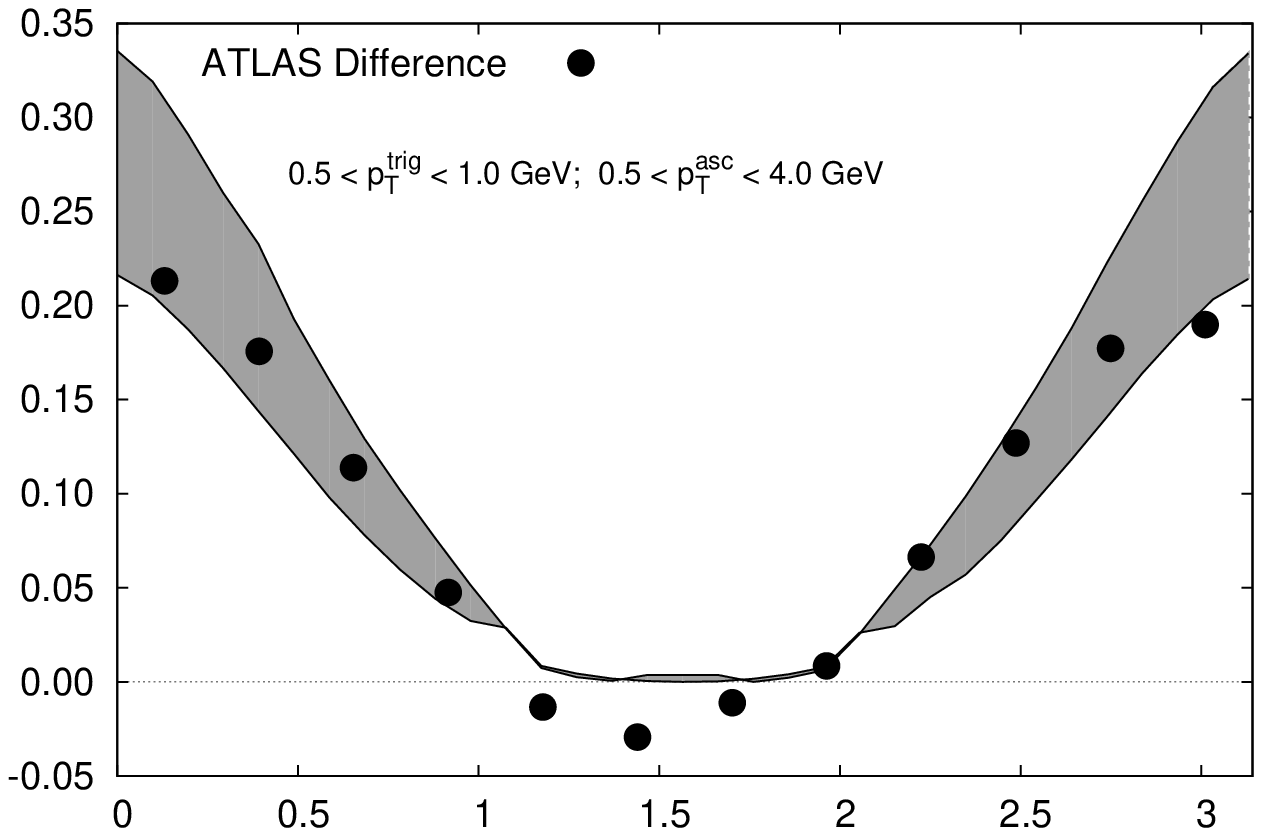}
\includegraphics[width=2.4in]{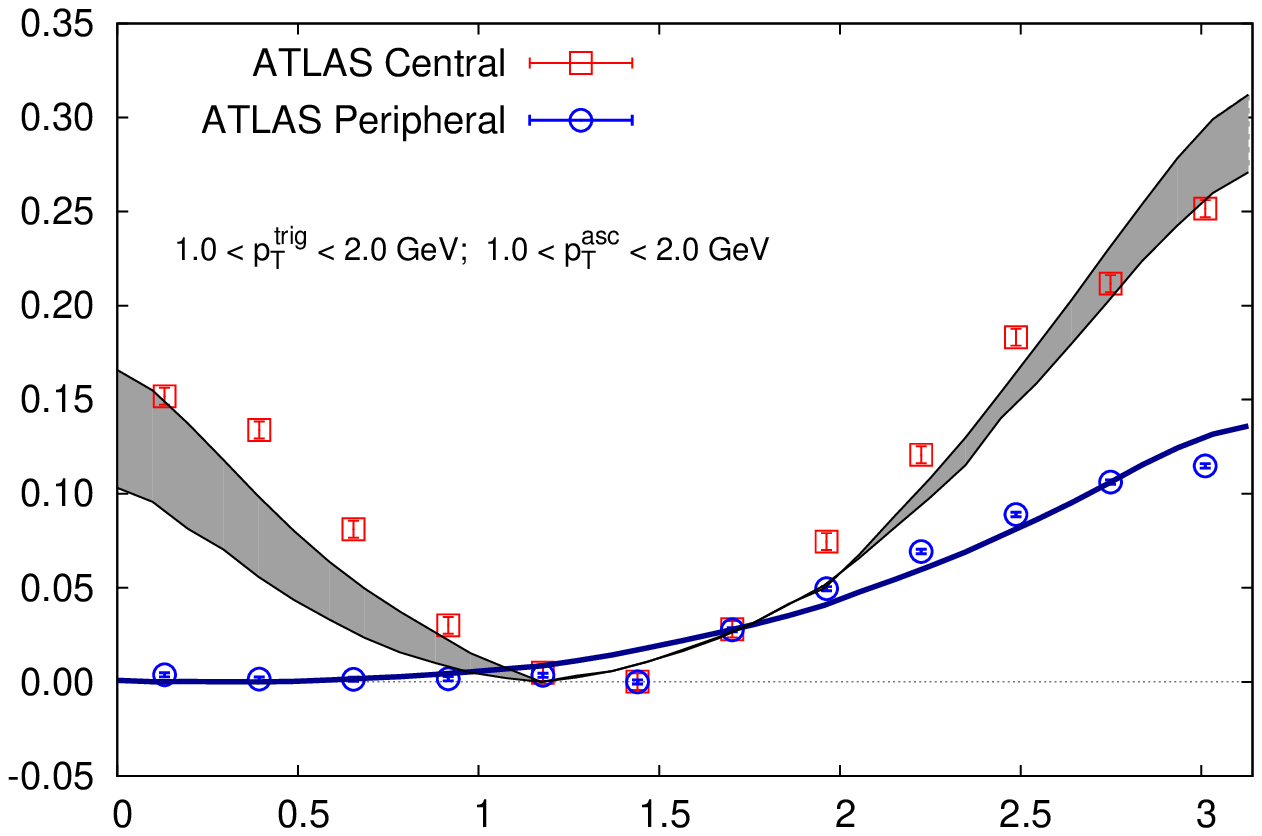}
\includegraphics[width=2.4in]{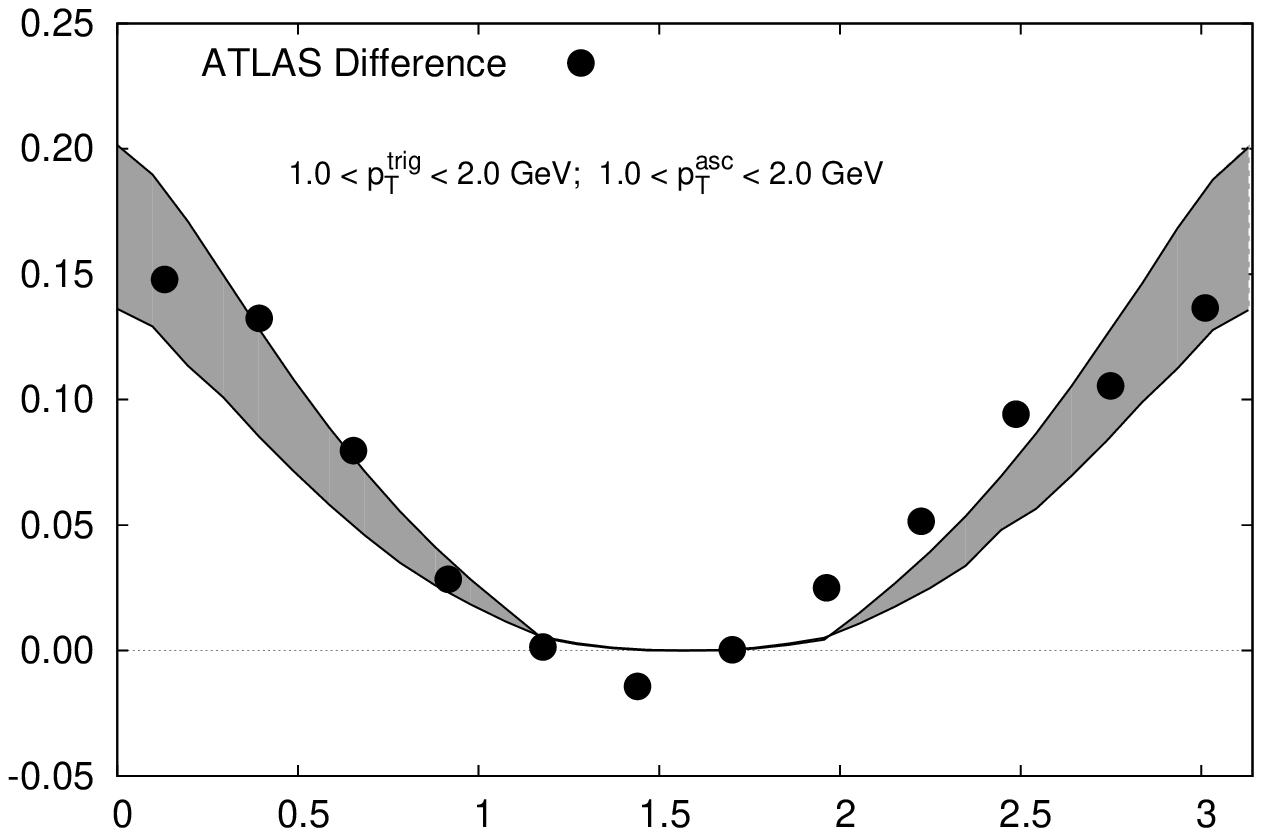}
\includegraphics[width=2.4in]{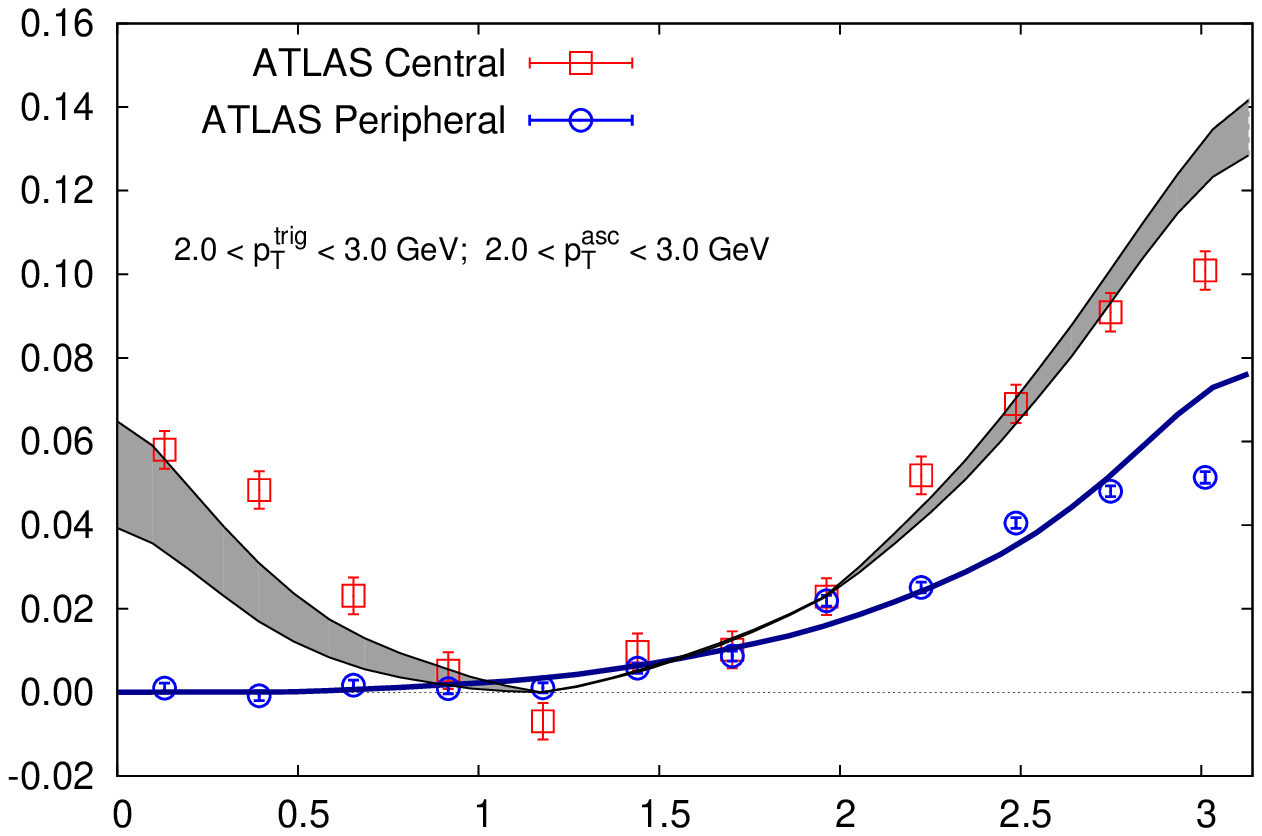}
\includegraphics[width=2.4in]{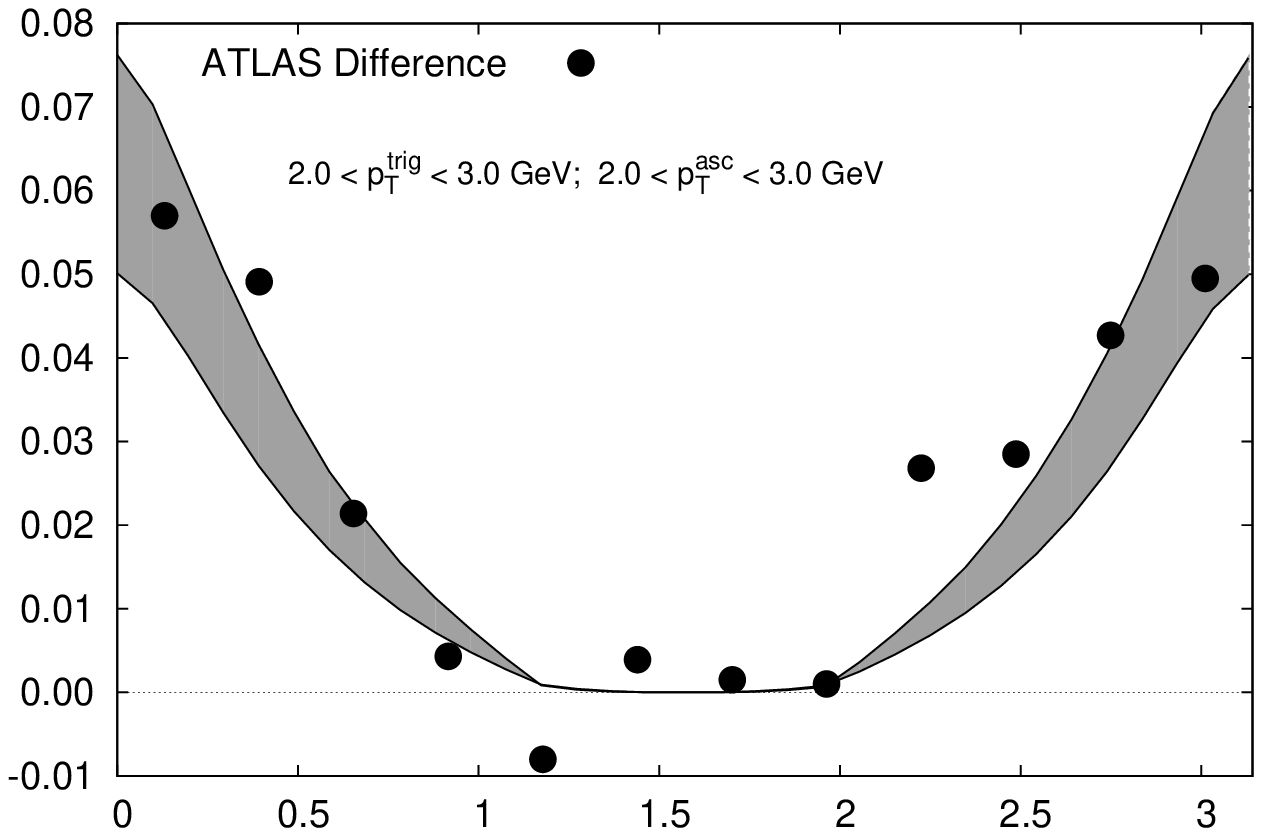}
\caption{Left: Data from the ATLAS collaboration for the associated yield versus $\Delta \phi$ for the central and peripheral 
events identified in~\cite{Aad:2012gla}. The blue theory curve corresponds to the comparison of Eq.~ (\ref{eq:BFKL}) for the min-bias $Q_0^2 (proton) = 0.168$ GeV$^2$ and $\Npb=3$ to the peripheral collision data. The other theory curve is a comparison of Eqs.~(\ref{eq:Glasma-corr})+ (\ref{eq:BFKL}) to the central data. The band corresponds to initial saturation scales, in our notation, of 
(4,14) (lower curve) and (3,22) (upper curve). Right: direct comparison of the central minus peripheral subtraction performed by ATLAS 
to the Glasma graph~(\ref{eq:Glasma-corr}) contribution.
}
\label{fig:atlas}
\end{figure}
%%%%%%%%%%%%%%%%%%%%%%%%%%%%%%%%%%%%%%%%%%%%%%%%%%%%%%%%%%%%%%%%%%%%%%%%%

%%%%%%%%%%%%%%%%%%%%%%%%%%%%%%%%%%%%%%%%%%%%%%%%%%%%%%%%%%%%%%%%%%%%%%%%%
\begin{figure}
\includegraphics[width=2.0in]{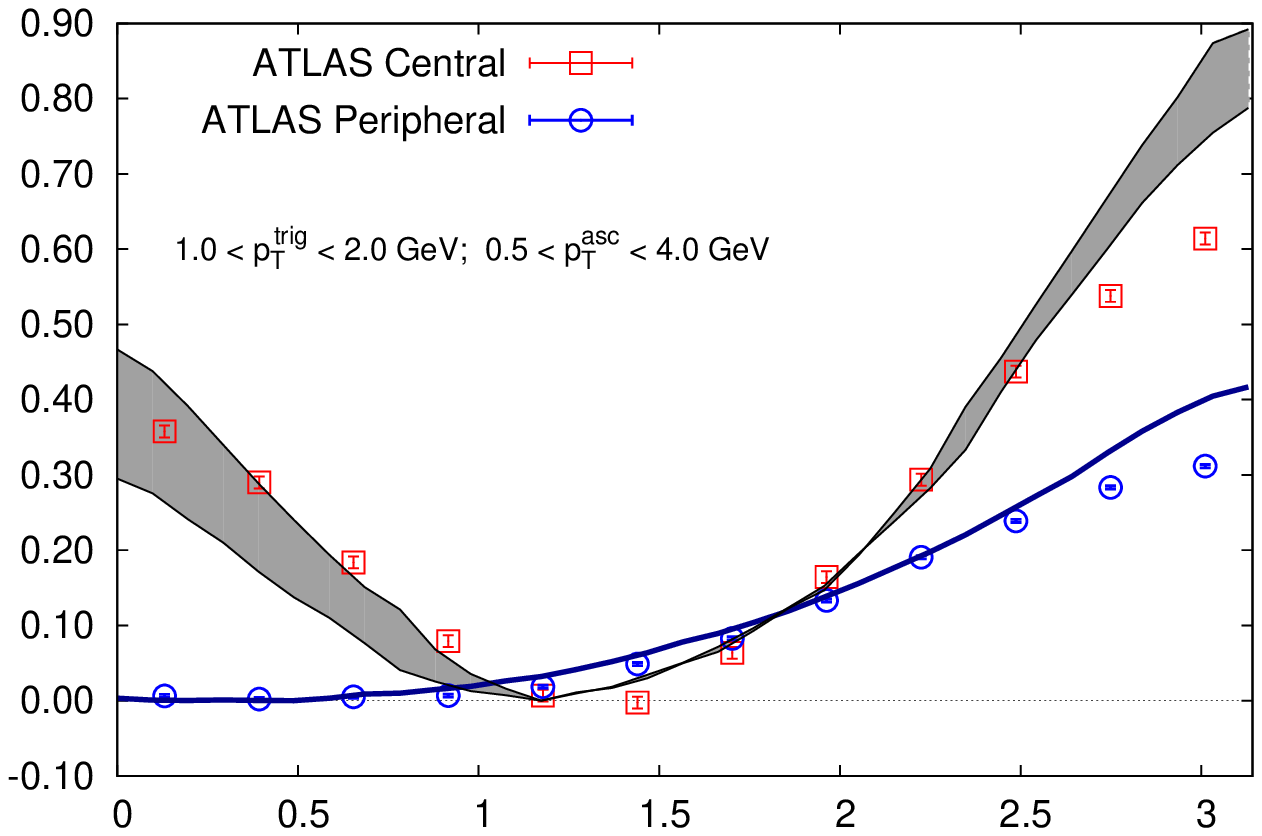}
\includegraphics[width=2.0in]{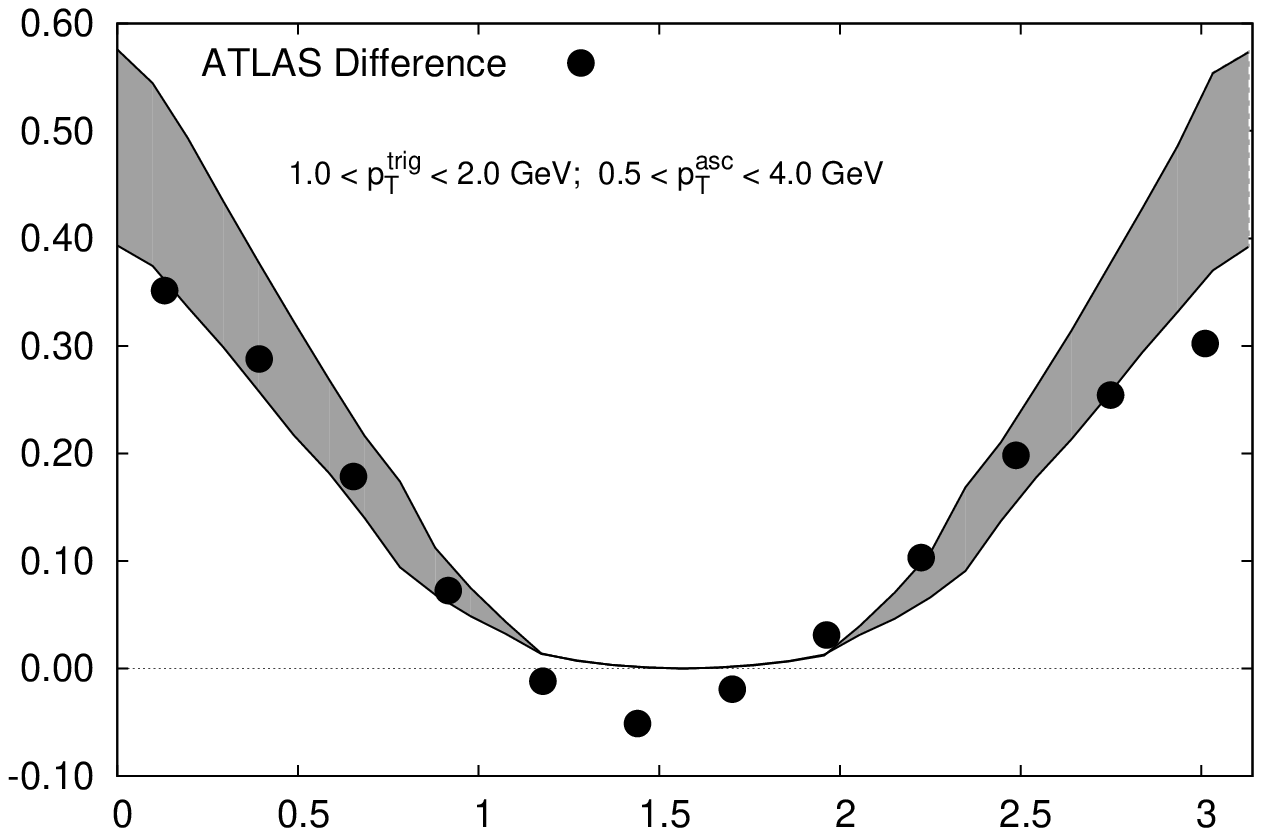}\\
\includegraphics[width=2.0in]{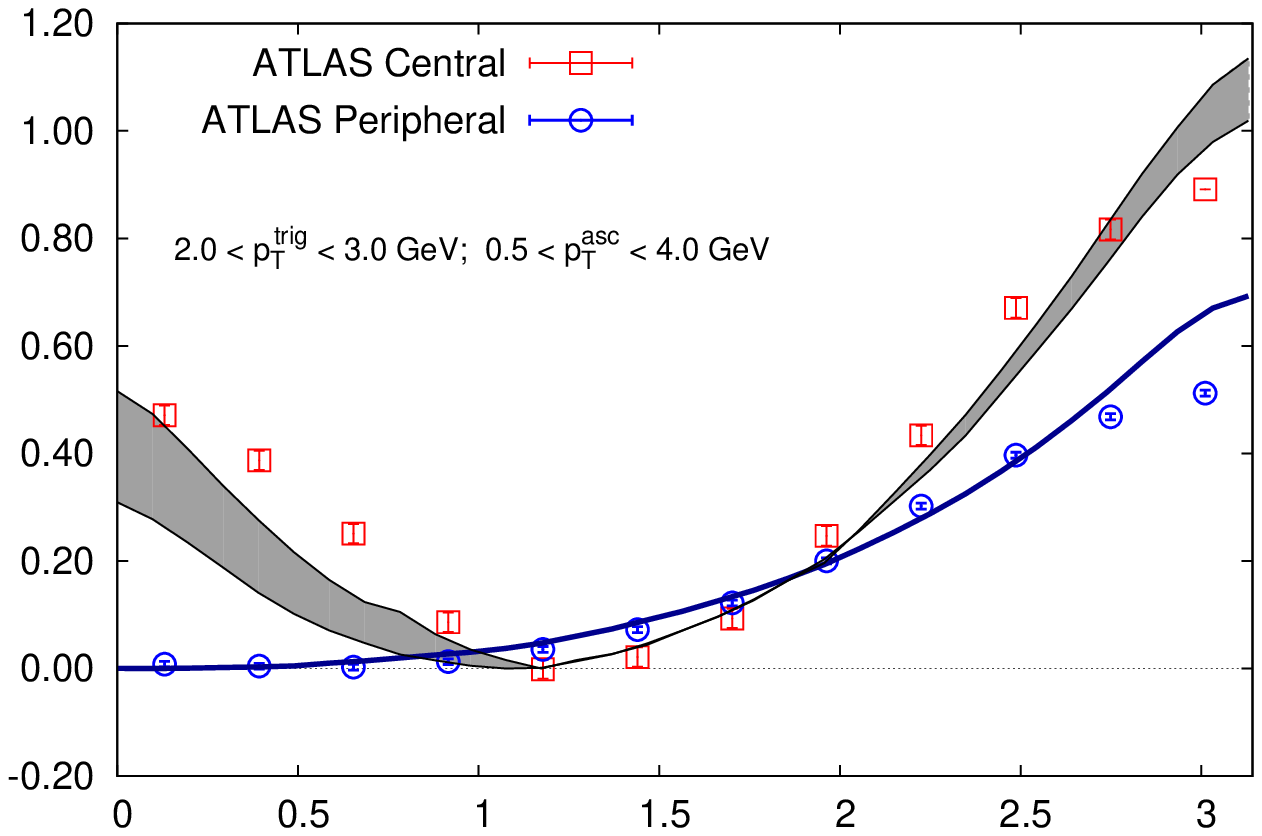}
\includegraphics[width=2.0in]{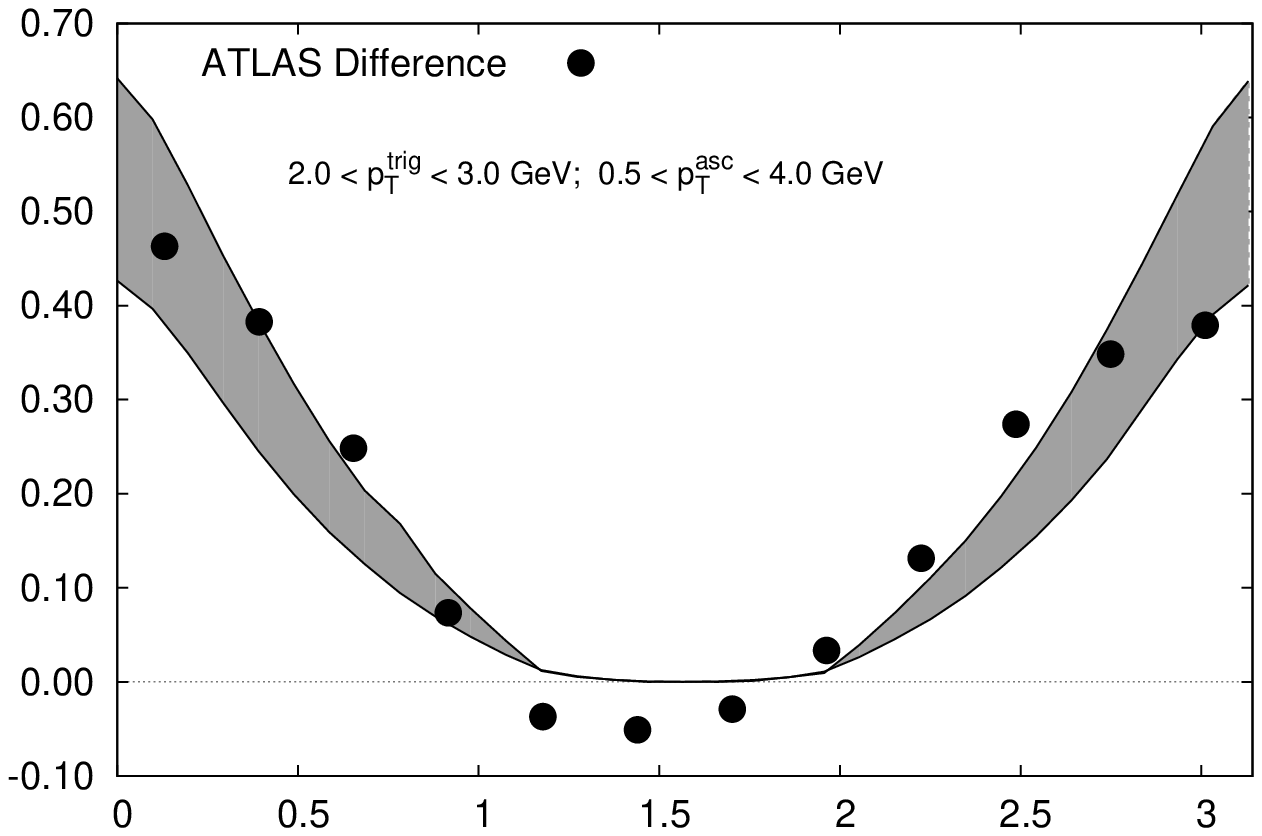}\\
\includegraphics[width=2.0in]{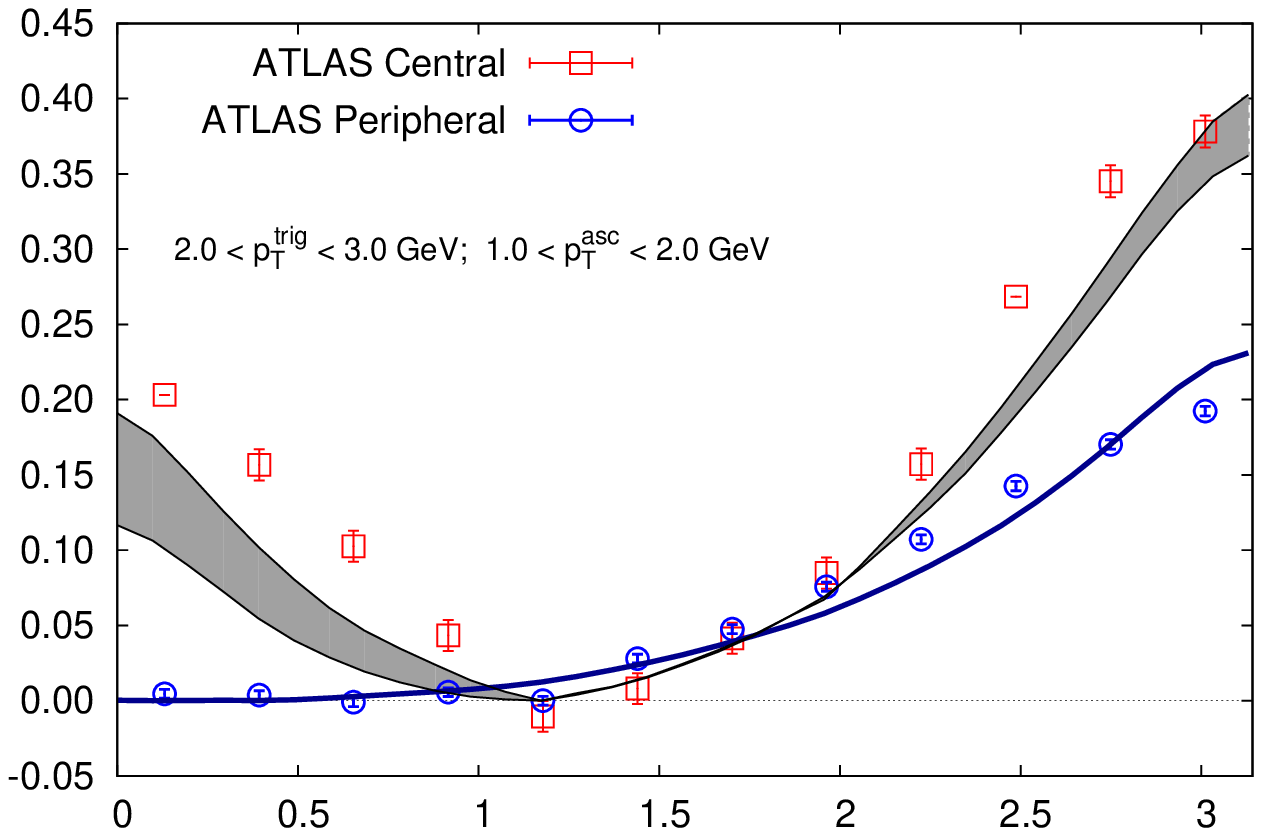}
\includegraphics[width=2.0in]{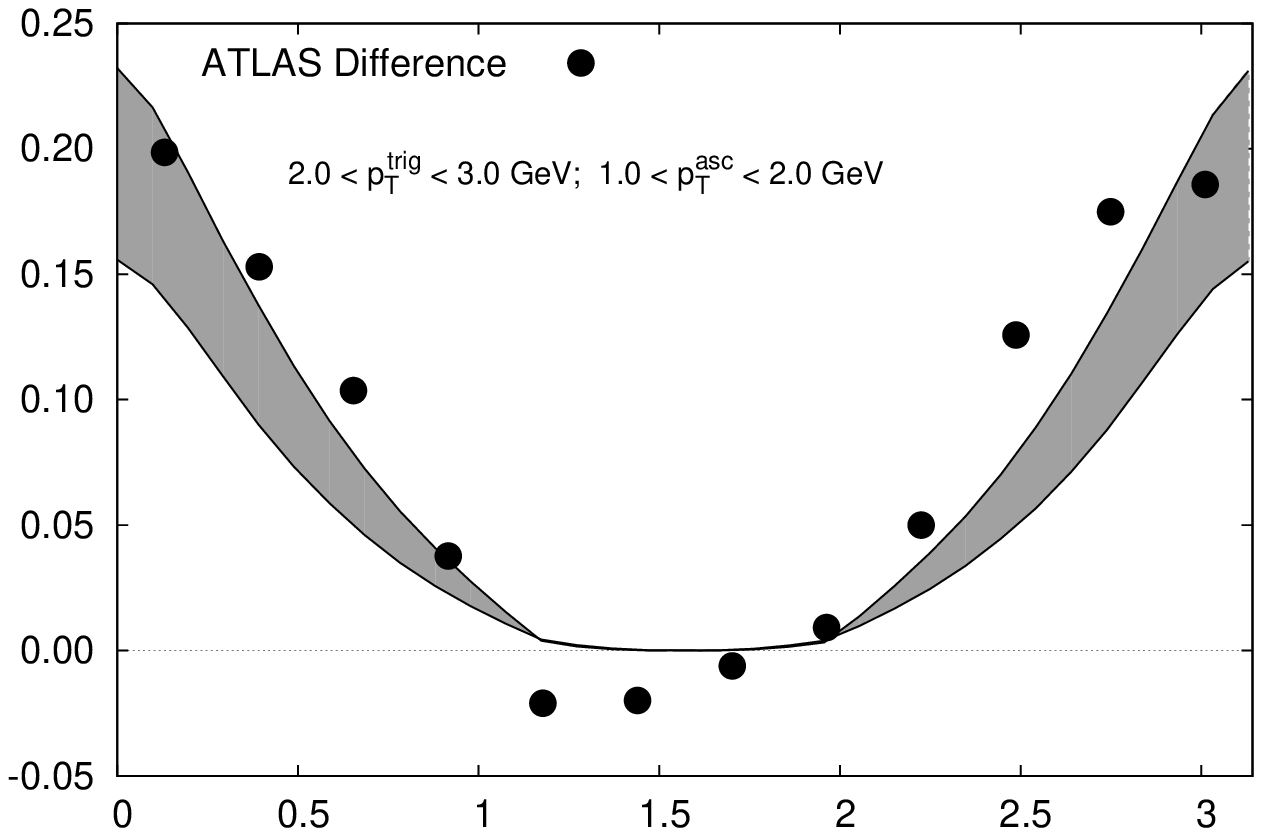}\\
\includegraphics[width=2.0in]{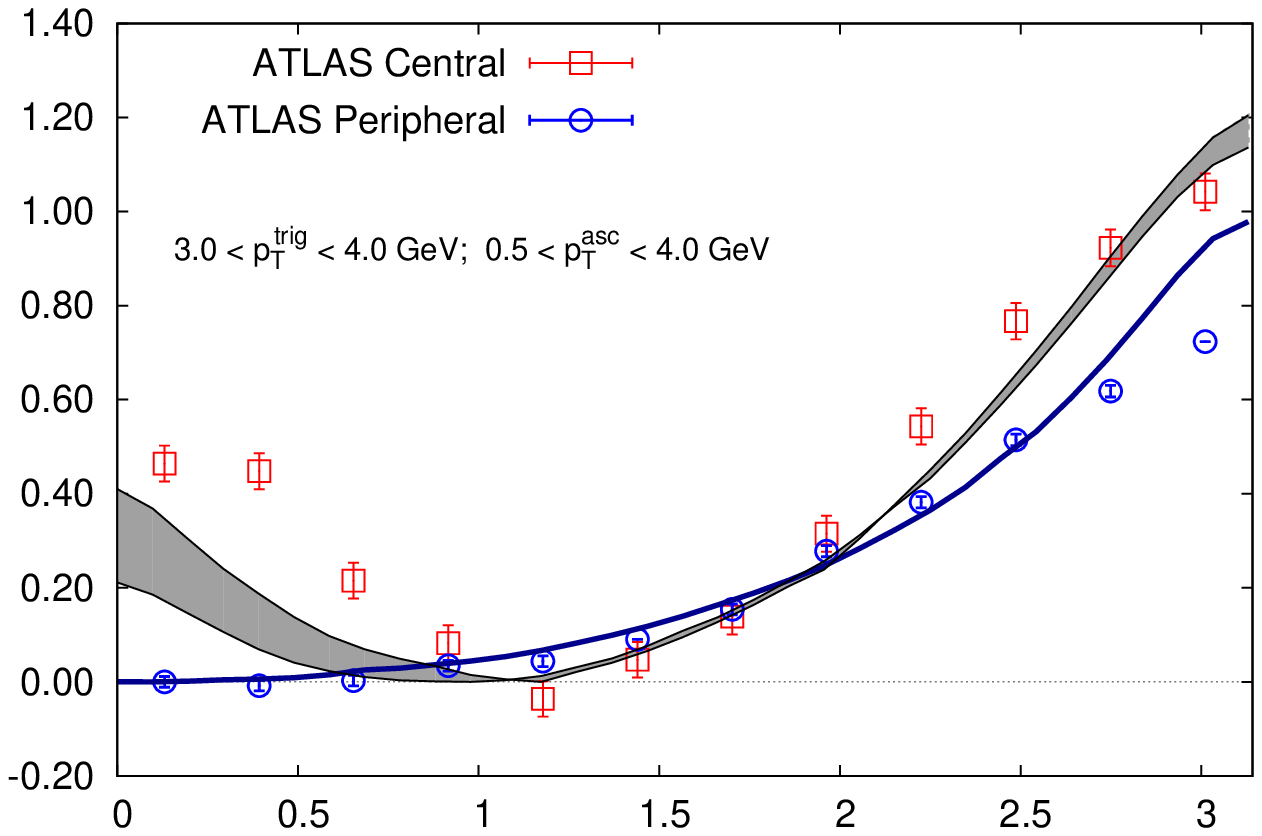}
\includegraphics[width=2.0in]{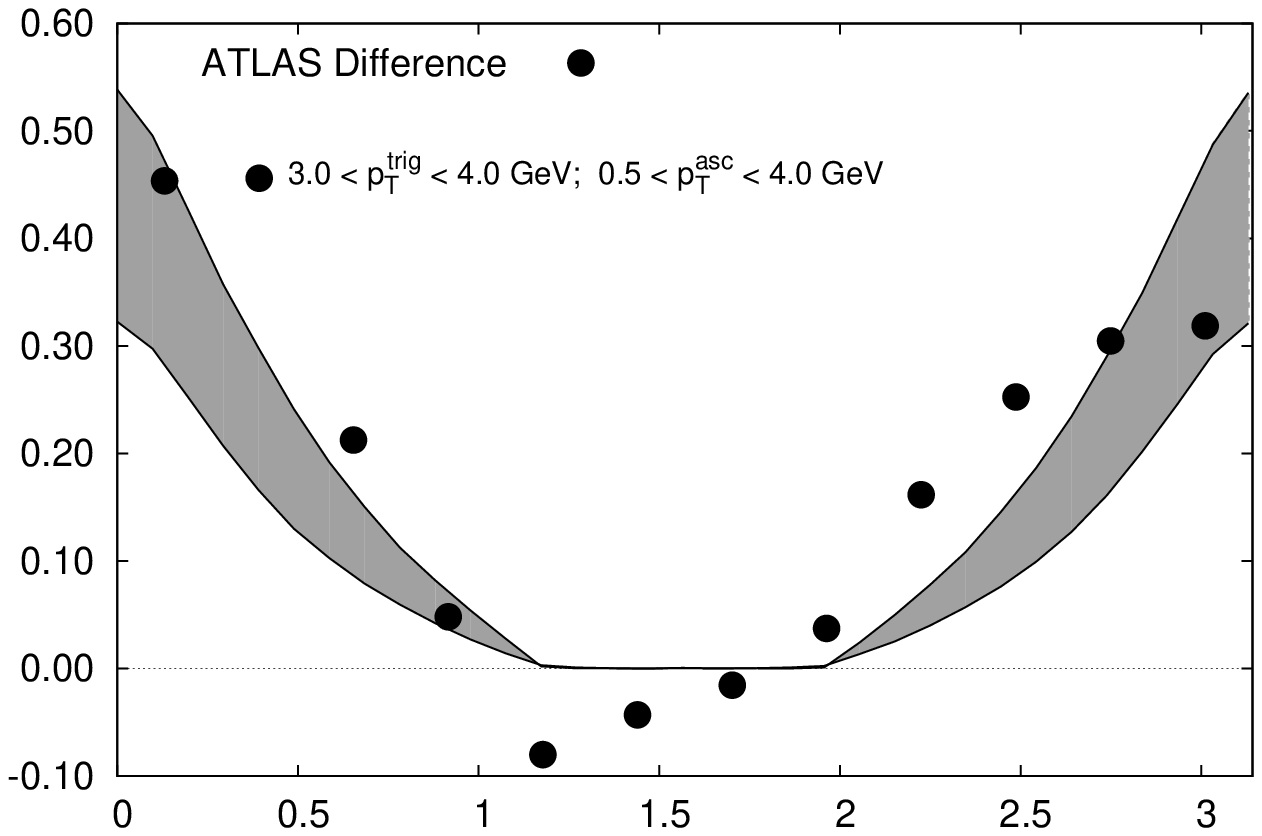}\\
\includegraphics[width=2.0in]{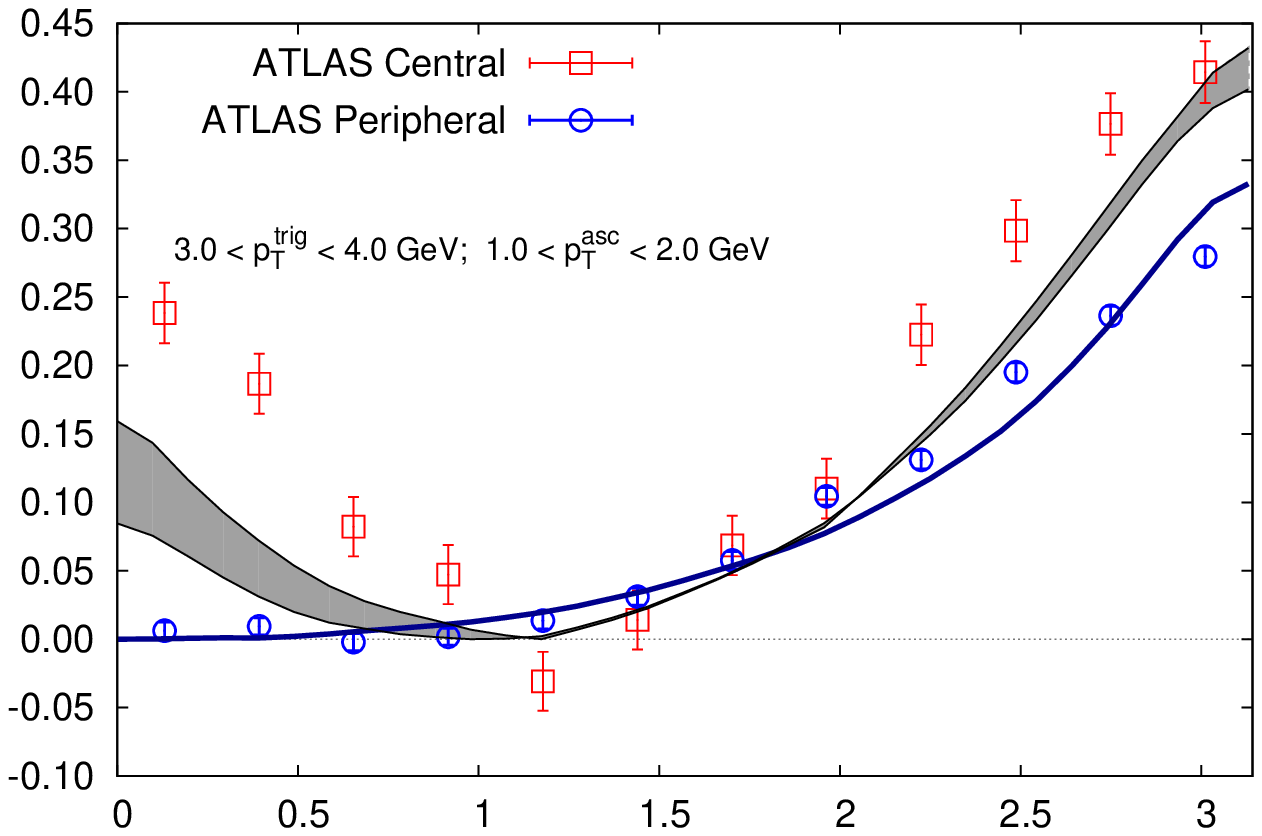}
\includegraphics[width=2.0in]{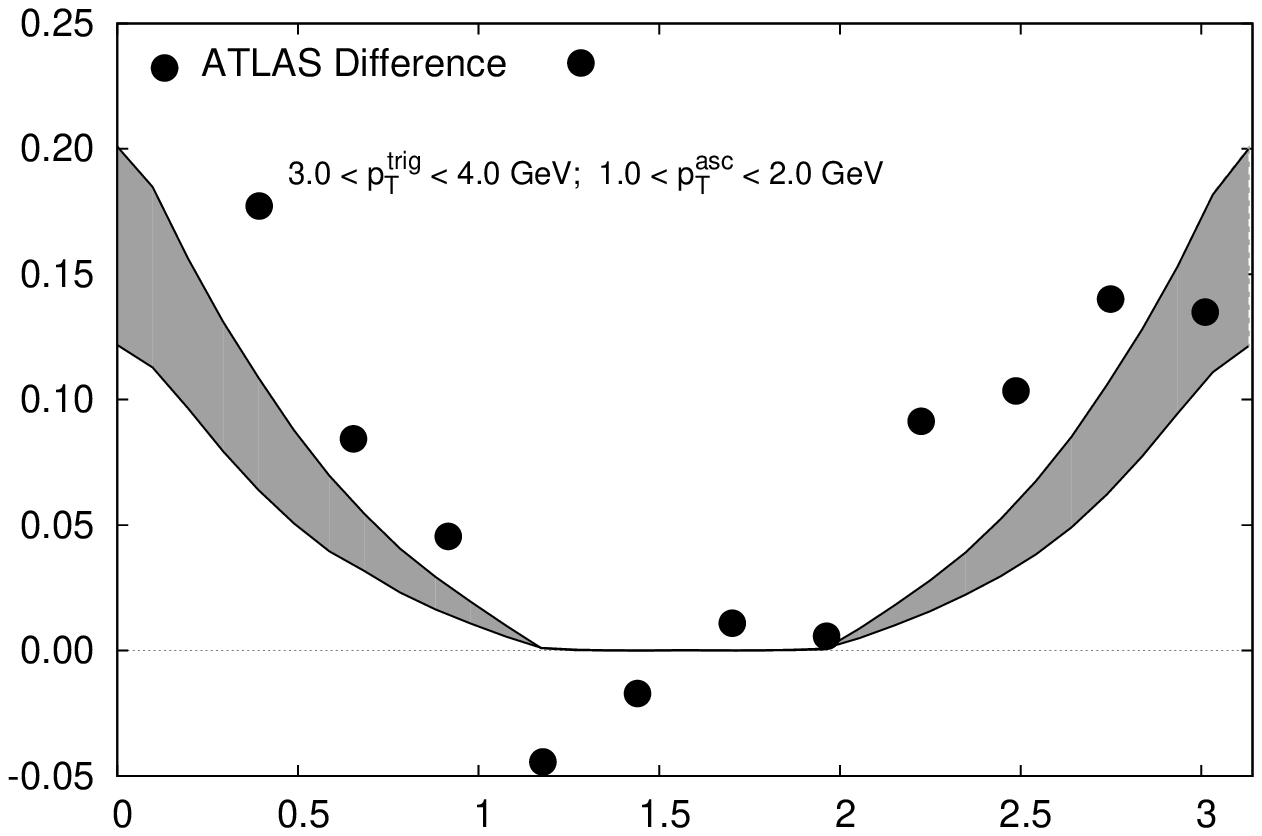}\\
\includegraphics[width=2.0in]{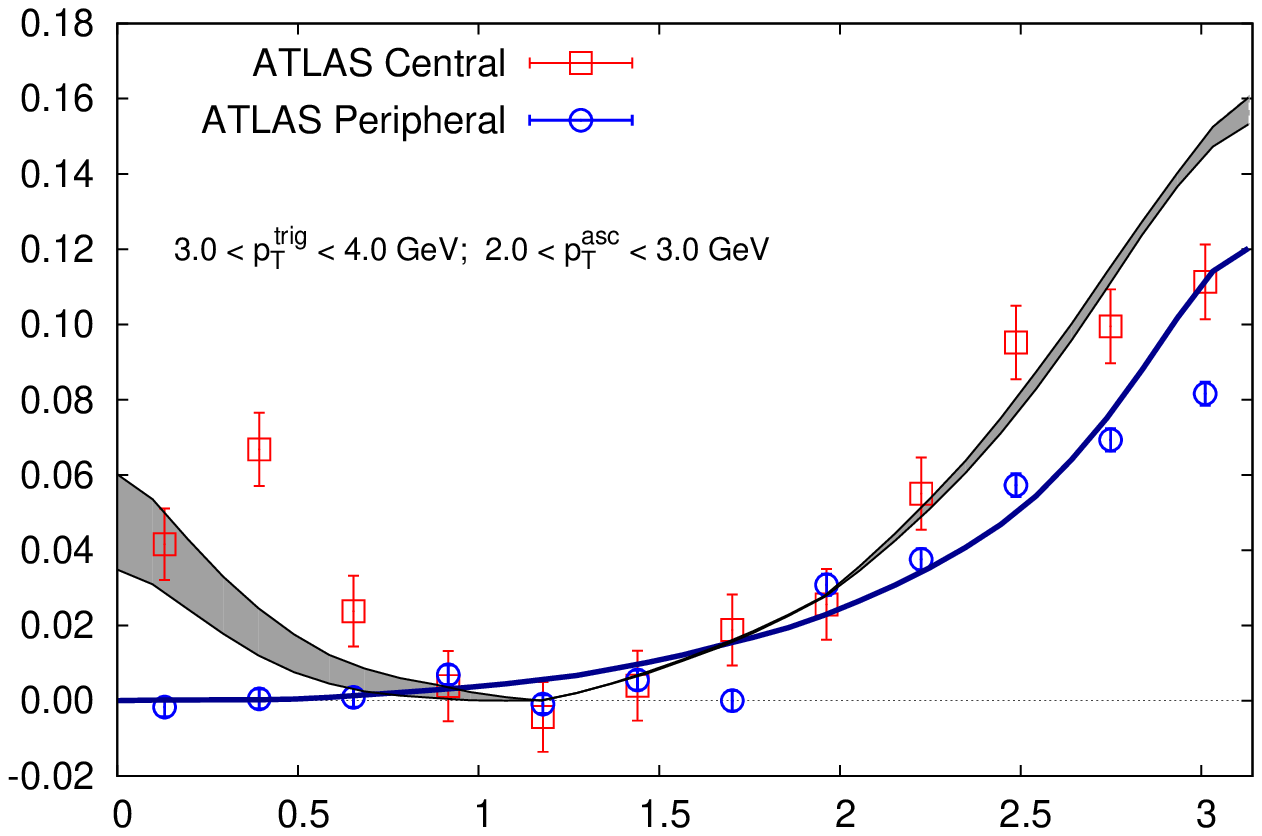}
\includegraphics[width=2.0in]{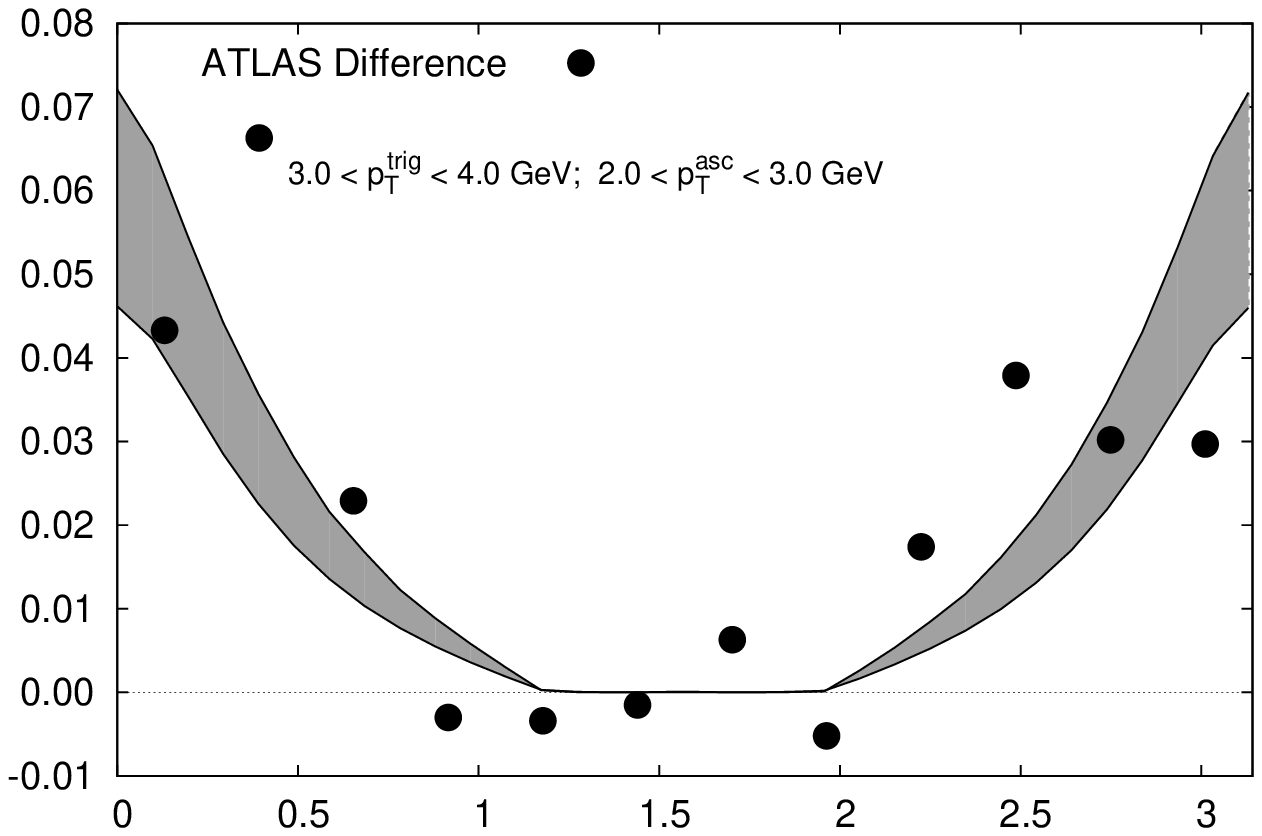}
\caption{Per-trigger-yield $1/N_{\rm trig} d^2N/d\Delta\phi$ versus $\vert \Delta\phi\vert$ for asymmetric $\ptrig$, $\ptasc$.  Red squares are the ATLAS data for central p+Pb events,  blue circles are for peripheral p+Pb events. The blue curves show the results of the  BFKL contribution for ($\Npp,\Npb$)=(1,3). 
The gray band, representative of central collisions, corresponds to (4,14)--lower curve and (3,22)--upper curve.}
\label{fig:atlas2}
\end{figure}
%%%%%%%%%%%%%%%%%%%%%%%%%%%%%%%%%%%%%%%%%%%%%%%%%%%%%%%%%%%%%%%%%%%%%%%%%

Recently, subsequent to our paper \cite{Dusling:2012wy}, papers on long range rapidity correlations in the LHC proton-lead data 
at $\sqrt{s}=5.02$ GeV from both ALICE~\cite{Abelev:2012ola} and
ATLAS~\cite{Aad:2012gla} collaborations have appeared. The former 
took data on very central events in the peudorapidity window $|\Delta \eta| < 1.8$, while the later has a pseudorapidity acceptance 
$2 < |\Delta \eta|< 5$, closer to the CMS acceptance. (For a detailed discussion of the different acceptances and normalization procedures of the different experiments, we refer the reader to the appendix.) A novel feature first introduced by ALICE, and shortly thereafter by ATLAS, is the subtraction of the associated yield per trigger in peripheral proton-nucleus from the same quantity in more central collisions. In our framework of Glasma graphs + BFKL graphs independently contributing to the yield per trigger, this procedure is especially valuable because it is equivalent to isolating the Glasma graph component \footnote{In contrast, such a procedure is not useful for an interpretation in a final state interaction scenario because this procedure implicitly assumes that di-hadrons of a given $p_T$ are not strongly interacting, while final state scenarios assume instead that they are strongly interacting. \label{ft:c1}}. As we checked previously, the BFKL di-jet per trigger contribution is very weakly centrality dependent~\cite{Dusling:2012cg}. 

In Fig.~(\ref{fig:atlas}),  we show a comparison of results in our framework to the ATLAS data in symmetric $p_T$ windows.  The two data sets at higher $p_T$ values are also available from the CMS 
experiment; as noted, ATLAS has a larger $\Delta \eta$ acceptance relative to CMS. In the left plot, we show a comparison of 
the ATLAS peripheral data to the BFKL contribution+Glasma contribution. The Glasma contribution is negligible in peripheral events. We see that $\Npp =1$ on the proton side and $\Npb=3$ on the lead side give a good description of the data. This adds confidence to our fits to the CMS data since the  quoted values for $Q_0^2$ are indeed of the order of what one would expect as typical values in peripheral collisions. The central ATLAS data should be compared to the sum of 
Glasma + BFKL contributions, and we see again we get good fits for the band (3,22) (upper curve), (4,14) (lower curve), as in the 
CMS comparison. The agreement is quite reasonable though the theory curves slightly under-predict the data in the plots of the net associated yield on the nearside (left plots of Fig.~(\ref{fig:atlas})).  We attribute this to an artificial anti-collimation of the BFKL contribution at small $\Delta\phi$.  After the jet subtraction performed by ATLAS the agreement between the Glasma Graphs and {\em ridge} yield is restored as demonstrated in the right plots of Fig.~(\ref{fig:atlas}), where the ``di-jet" subtracted yield is compared to our Glasma graph. The overall 
agreement is quite reasonable suggesting a consistent interpretation of the power counting in our ``initial-state" framework for multi-particle production and the experimental observations. 

The ATLAS collaboration has presented data for proton-lead collisions in a wide range of $\ptrig, \ptasc$ windows for central and peripheral collisions. In addition to the symmetric $p_T=\ptrig=\ptasc$ windows have already been presented by CMS for a similar acceptance, we consider in Fig.~(\ref{fig:atlas2}) a comparison to asymmetric $\ptrig\neq\ptasc$ windows. We have also included in our comparison the 
data for $\ptasc$, where the lower range of the window is $p_T=0.5$ GeV, which one might consider quite low $p_T$'s for our framework. For the asymmetric windows shown in Fig.~(\ref{fig:atlas2}), the agreement on the awayside is quite good for both central and peripheral events, though arguably underpredicting the central events. However, on the nearside, the Glasma contribution is 
significantly lower than the data for central events, especially for the larger $\ptrig$ windows. Of all the $\ptrig,\ptasc$ windows 
presented by ATLAS, the asymmetric $3<\ptrig<4$ GeV windows provide the worst comparison to the nearside Glasma computation. On the theory side, we have not attempted any fine tuning with small systematic adjustments of the 
$K$ factors and $Q_0^2$ values.  At these higher $p_T$ values the same trigger bias, as discussed earlier, has to be taken into consideration.  It will be interesting to see if this discrepancy persists with the additional data that is anticipated to be released soon, and if this discrepancy in asymmetric windows is also seen by CMS and ALICE. 

%%%%%%%%%%%%%%%%%%%%%%%%%%%%%%%%%%%%%%%%%%%%%%%%%%%%%%%%%%%%%%%%%%%%%%%%
\begin{figure}[]
\includegraphics[width=0.5\textwidth]{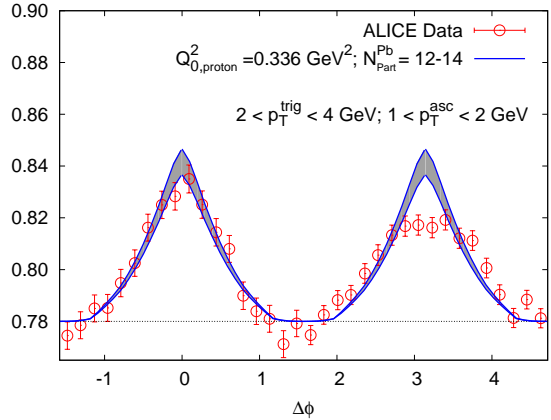}
\caption{Comparison of Glasma graphs, Eq.~(\ref{eq:Glasma-corr}), to ALICE data for the collimated yield per trigger per $\Delta \eta$ 
for central (0-20\%) events with (60-100\%) peripheral events subtracted versus $\Delta \phi$. The band is from a computation with $Q_0^2$(proton) = 0.336 GeV$^2$ on $N_{\rm part}^{\rm Pb}=12$ (lower) and $N_{\rm part}^{\rm Pb}=14$ (upper).}
\label{fig:alice1}
\end{figure}
%%%%%%%%%%%%%%%%%%%%%%%%%%%%%%%%%%%%%%%%%%%%%%%%%%%%%%%%%%%%%%%%%%%%%%%%%

%%%%%%%%%%%%%%%%%%%%%%%%%%%%%%%%%%%%%%%%%%%%%%%%%%%%%%%%%%%%%%%%%%%%%%%%
\begin{figure}[]
\includegraphics[width=0.5\textwidth]{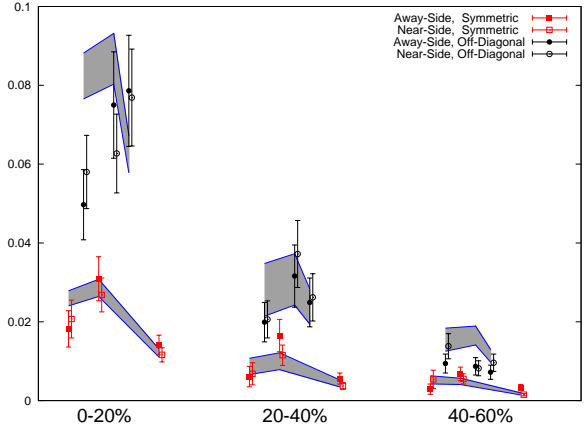}
\caption{Comparison of the (60-100\% centrality subtracted) data for both nearside and awayside integrated collimated yields from different centrality classes for ``off-diagonal"  windows ($2<\ptrig<4$ GeV: i) $0.5<\ptasc<1$ GeV, ii) $1<\ptasc < 2$ GeV, 
and $1<\ptrig<2$ GeV, $0.5<\ptasc<1$ Gev) and symmetric windows ($\ptrig=\ptasc =0.5-1, 1-2, 2-4$ GeV). 
The band for the 0-20\% centrality class is the result for (2,12) (lower curve) and (2,14) (upper curve), 
20-40\% centraliy: (2,4) (lower), (2,6) (upper) and  40-60\% centrality: (1,3) (lower), (1,4) (upper).}
\label{fig:alice2}
\end{figure}
%%%%%%%%%%%%%%%%%%%%%%%%%%%%%%%%%%%%%%%%%%%%%%%%%%%%%%%%%%%%%%%%%%%%%%%%%

Data on proton-lead collisions from the ALICE collaboration is shown in
Figs.~(\ref{fig:alice1}) and (\ref{fig:alice2}). The acceptance of the ALICE
experiment is distinctly different from CMS or ATLAS, covering $|\Delta \eta| <
1.8$. For small $|\Delta \eta| < 1$, there is a 
nearside $\Delta\phi\approx 0$ di-hadron correlation that can be attributed to
jet fragmentation. If this short range component can be safely subtracted, one
can look for a contribution that is long range in rapidity. Further subtraction
of the awayside ``di-jet" contribution, whose yield per trigger is observed to
be weakly centrality dependent, will, as discussed, reveal the Glasma graph
contribution. The Glasma graph contribution is nearly rapidity independent (for
$|\Delta \eta| \sim 1/\alpha_S$~5 units of $\Delta \eta$), so one expects such a
contribution to be present in the ALICE acceptance. In Fig.~(\ref{fig:alice1}), we show the comparison of the Glasma graph computation to the central ($0-20\%$) minus peripheral ($60-100\%$) yield from ALICE. The agreement is quite good. The overshoot on the awayside is sensitive to smearing of the back-to-back contribution and possible systematic uncertainties due to the peripheral jet subtraction. 

We now turn to the description of the integrated nearside and awayside ALICE
data in Fig.~(\ref{fig:alice2}). We see that for nearly all the symmetric and
asymmetric, near and awayside $p_T$ windows, and all centrality classes listed
in the caption of Fig.~(\ref{fig:alice2}), there is good agreement with the
ALICE data. There is a slight overshoot of the data in the lowest $p_T$ window
for the most central collisions and for one of the windows in the 40-60\%
centrality range. The ALICE paper~\cite{Abelev:2012ola} also quotes values for the $v_2$ and $v_3$ flow moments. In our framework, extraction of these moments also depends on the combinatorial $\Delta\phi$ independent background, which varies from one $p_T$ window to the next. Since there are many possible QCD contributions to the combinatorial background~\cite{Dumitru:2010mv,Dumitru:2011zz,Kovner:2010xk,Kovner:2011pe}, an estimate of this quantity is less reliable in our framework, though in principle feasible in future. For the connected graphs we have 
considered so far, there can be no $v_3$ contribution because these Glasma contributions are symmetric about $\Delta\phi=\pi/2$. It remains an open question whether the full set of connected graphs can produce a small $v_3$ collimation due to weak final state effects. In any event, a large $v_3$ component would be challenging for our framework. Our take on the ALICE $v_3$ data presented in 
~\cite{Abelev:2012ola} is that the effect observed is sensitive to systematics of the jet subtraction--see also the comment in footnote~\ref{ft:c1}. Further data from the 2013 p+Pb run should help clarify these issues significantly. 

%%%%%%%%%%%%%%%%%%%%%%%%%%%%%%%%%%%%%%%%%%%%%%%%%%%%%%%%%%%%%%%%%%%%%%%%%
\begin{figure}
\includegraphics[width=3.2in]{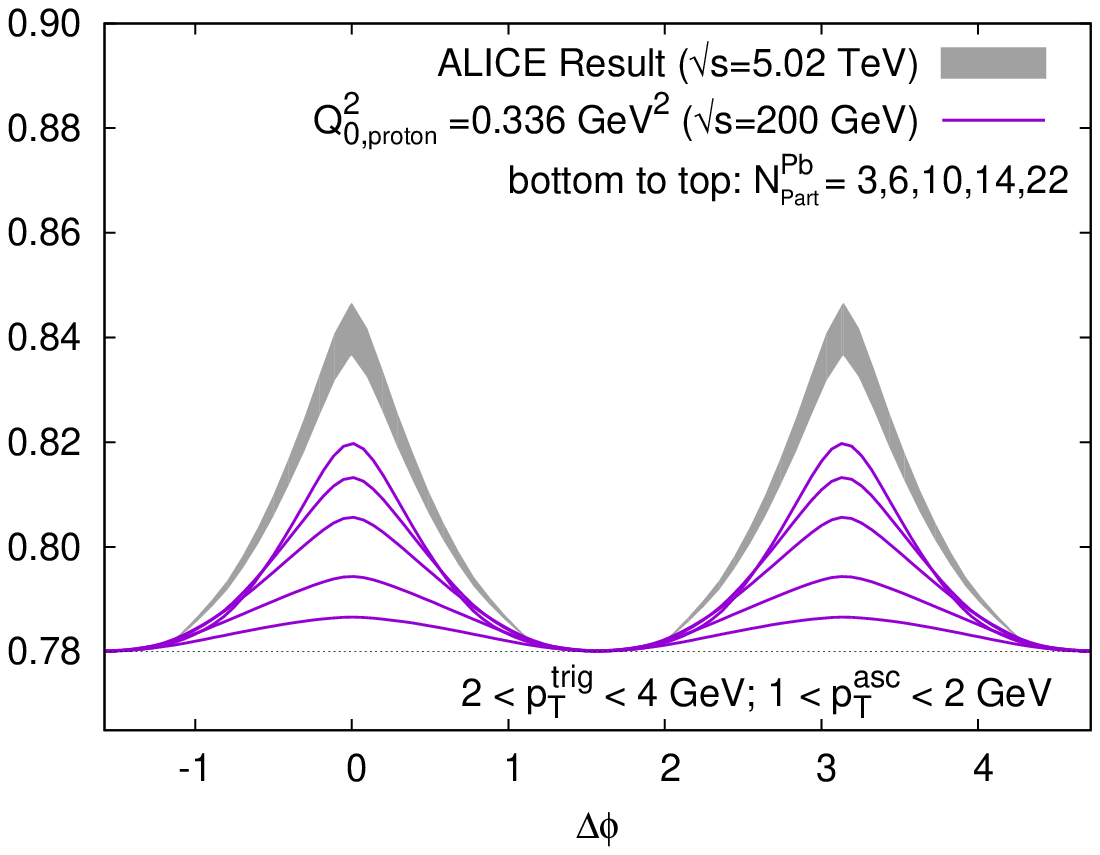}
\includegraphics[width=3.2in]{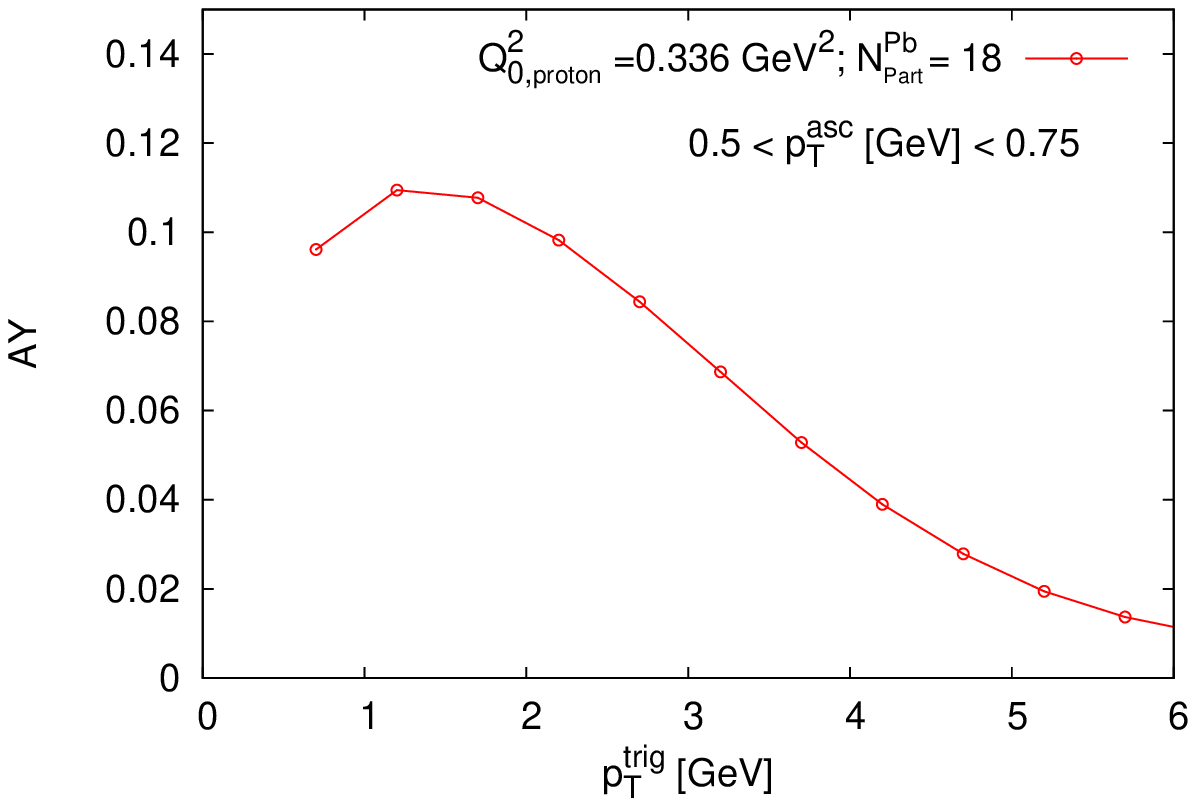}
\caption{Predictions for the associated yield versus $\Delta \phi$ (left figure) and the $\Delta \phi$ integrated associated yield $\ptrig$ (right) figure from Glasma graphs for RHIC energies. The result from Fig.~\ref{fig:alice2} for the ALICE experiment is show for comparison.}
\label{fig:phenix}
\end{figure}
%%%%%%%%%%%%%%%%%%%%%%%%%%%%%%%%%%%%%%%%%%%%%%%%%%%%%%%%%%%%%%%%%%%%%%%%%

Last but not least, in Fig.~(\ref{fig:phenix}), we show predictions for the collimated yield from a correlated di-hadron signal that is 
long range in $\Delta \eta$ for RHIC energies. It is clear from Fig.~(\ref{fig:multi_revised}) that a collimated signal is feasible at the significantly lower energies of deuteron-gold collisions at RHIC as long as very central events are triggered on. The predictions for different centrality classes and the functional dependence of the collimated yield on $\ptrig$ are shown in Fig.~(\ref{fig:phenix})--the ALICE prediction is shown in comparision. A cautionary aspect of comparisons to RHIC energies is that the higher $p_T$ windows and  $\Delta \eta > 0$ correlations are sensitive to $x > 0.01$ values in the nucleon and nuclear wavefunctions. The CGC framework should break down at these large $x$ values. For modest $\ptrig,\ptasc$ windows at RHIC, where the physics is still weak coupling and $x$ is still small, one expects important information due to the widely different energy to help constrain our picture of long range rapidity correlations. Very recently, first data from these correlations in deuteron-gold collisions at $\sqrt{s}=200$ GeV from the PHENIX experiment were presented~\cite{Sickles-WWND}. The analysis was very similar to that of the ATLAS experiment. A significant $v_2$ is observed, while $v_3$ is consistent with zero.  As we noted previous, due to the sensitivity of the result to the combinatorial background, a direct comparison of results similar to Fig.~(\ref{fig:phenix}) to the data cannot yet be achieved. We expect that results for the collimated associated yield per trigger will become available soon facilitating a more direct comparison to our predictions for RHIC energies.

\section{Summary and Outlook}

In this third paper in the series, preceded by \cite{Dusling:2012cg,Dusling:2012wy}, we developed significantly the comparison of the 
CGC EFT to large $\Delta \eta$ di-hadron correlation data from the LHC
experiments. The different normalizations of data taken by the different
experiments were taken into account and a consistent treatment of CMS, ALICE and
ATLAS data was presented -- the last two being discussed for the first time. In the CMS case, we presented a reanalysis relative to ~\cite{Dusling:2012cg,Dusling:2012wy} both for the proton-proton and proton-nucleus data. In the former case, we showed for the first time detailed comparisons of theory to data from a number of $\Ntrk$ windows. 

The agreement of theory with data in proton-proton and proton-lead collisions over a
 very wide range of $\ptrig,\ptasc$ windows, centrality class and $\Delta\eta$ acceptance, is quite spectacular and lends strong 
support that a) gluon saturation is being seen in these experiments, and b) the data are sensitive to systematics of renormalization group evolution of unintegrated gluon distributions (UGD) in the description of Glasma and ``back-to-back" graphs. It is remarkable that gluons widely separated in rapidity show a $\Delta \phi \approx 0$ collimation that is sensitive to detailed dynamical features of 
the theory. Because the Glasma graphs are quantum interference graphs, which have different structures in the amplitude and complex conjugate amplitude, the result is a particular form of gluon entanglement, unique to QCD. If this picture is confirmed by further data, it provides an impetus for further developments in theory to better understand the properties of saturated gluon states in the nucleon and nuclear wavefunctions, providing an important window to hadron structure and dynamics at high energies. 

An alternative scenario for the dynamics of long range rapidity correlations is 
from flow resulting from strong final state
rescattering~\cite{Bozek:2011if,Bozek:2012gr,Bozek:2013df}.  We believe this
possibility is strongly disfavored on both conceptual and phenomenological
grounds--these issues will be addressed elsewhere. Our perspective based on the
discussion in this paper is that the onus is now on models based on final state
scenarios to present a quantitative description of data that is as transparent and competitive to the same degree as the one presented here--for instance, the matrix of data of the collimated yield versus $\Delta \phi$ for 
the various experiments. 

Further data from the LHC will no doubt provide clarity and challenge either or both scenarios. Regardless, unless the data presented thus far changes significantly, the results of our detailed analysis suggest that initial state effects must play an important role even if other QCD effects come into consideration in the description of these striking experimental phenomena. 

\section*{Acknowledgements}
We are especially grateful to Jan Fiete Grosse-Oetringhaus, Constantin Loizides,
Jiangyong Jia, Wei Li, Gunther Roland and Anne Sickles for very valuable
discusssions on experimental issues. We would further like to thank Adam Bzdak,
Subhasis Chattopadhyay, Adrian Dumitru, Yuri Kovchegov, Roy Lacey, Larry McLerran, Bjoern Schenke, Prithwish Tribedy as well as Helen Caines, John Harris and members of their Yale Relativistic Heavy Ion Group for very useful input. K.D. and R.V are supported by the US Department of Energy under DOE Contract Nos. DE-FG02-03ER41260 and DE-AC02-98CH10886 respectively.  This research used resources of the National Energy Research Scientific Computing Center, which is supported by the Office of Science of the U.S. Department of Energy under Contract No. DE-AC02-05CH11231.

\section*{Appendix: Experimental coverage and relative normalizations of di-hadron data in the LHC experiments}

In this appendix, we discuss the relative experimental acceptances and normalizations of the different LHC experiments. The rapidity acceptance of these are 

\begin{center}
\vspace{1em}
\begin{tabular}{ l | c c c c }
& $\eta_{\rm min}$ & $\eta_{\rm max}$ & $\Delta\eta_{\rm min}$ & $\Delta\eta_{\rm max}$ \\
\hline
CMS & -2.4 & +2.4 & 2.0 & 4.0 \\
ALICE & -0.9 & +0.9 & 0 & 1.8 \\
ATLAS & -2.5 & +2.5 & 2.0 & 5.0\\
\end{tabular}
\vspace{1em}
\end{center}

The effect of different normalizations between the experiments can be  gauged by considering  single inclusive and double inclusive charged hadrons obtained from Eqs.~(\ref{eq:ntrig}) and (\ref{eq:dihadron}). In this case, 
\begin{equation}
\Ntrig= \int_{\eta_{\rm min}}^{\eta_{\rm max}} d\eta \int_{p_T^{\rm
min}}^{p_T^{\rm max}} d^2\pp \frac{dN_{\rm ch}}{d\eta d^2\pp} 
\label{eq:ntrig2}
\end{equation}
and 
\begin{equation}
\frac{d^2 N}{d\Delta\phi} = \int_{\eta_{\rm min}}^{\eta_{\rm max}} d\eta_p
\int_{\eta_{\rm min}}^{\eta_{\rm max}} d\eta_q \int_{p_T^{\rm min}}^{p_T^{\rm
max}} d^2\pp \int_{q_T^{\rm min}}^{q_T^{\rm max}} d^2\qp
\;\frac{\mathcal{A}\left(\eta_p,\eta_q,\phi_p,
\phi_q\right)}{\mathcal{B}\left(\eta_p,\eta_q\right)}\;\frac{d^2N_{\rm ch}}{d\eta_p d\eta_q d^2\pp d^2\qp}
\label{eq:d2N}
\end{equation}
where 
\begin{equation}
\mathcal{A}\left(\eta_p,\eta_q,\phi_p, \phi_q\right) =
\delta(\phi_p-\phi_q-\Delta\phi)\theta\left(\vert
\eta_p-\eta_q\vert-\Delta\eta_{\rm min}\right)\theta\left(\Delta\eta_{\rm
max}-\vert\eta_p-\eta_q\vert\right) \, .
\end{equation}

In Eq.~(\ref{eq:d2N}), the ATLAS experiment does not weight their signal by the background nor do they
calculate their result per unit $\Delta\eta$.  For ATLAS, we have therefore $\mathcal{B}=1$.  For CMS and ALICE,
\begin{eqnarray}
\mathcal{B}_{\rm CMS}=\mathcal{B}_{\rm ALICE}=2\vert\Delta\eta_{\rm max}-\Delta\eta_{\rm
min}\vert \left(1-\frac{\vert \eta_p-\eta_q\vert}{\vert
\eta_{\rm max} - \eta_{\rm min}\vert}\right)\nonumber\\
\end{eqnarray}

To get a sense of what these different normalizations entail, let us compute 
\begin{equation}
\tN \sim \int_{\eta_{\rm min}}^{\eta_{\rm max}} d\eta
\end{equation}
and 
\begin{equation}
\frac{d^2 {\tilde N}}{d\Delta\phi} \sim \int_{\eta_{\rm min}}^{\eta_{\rm max}} d\eta_p
\int_{\eta_{\rm min}}^{\eta_{\rm max}} d\eta_q
\;\frac{\mathcal{A}\left(\eta_p,\eta_q,\phi_p,
\phi_q\right)}{\mathcal{B}\left(\eta_p,\eta_q\right)} \, .
\end{equation}

One will then have 
\begin{eqnarray}
\left.\frac{1}{\tN}\frac{d^2 {\tilde N}}{d\Delta\phi}\right|_{\rm CMS}&\sim&1.0 \nonumber\\
\left.\frac{1}{\tN}\frac{d^2 {\tilde N}}{d\Delta\phi}\right|_{\rm ALICE}&\sim&1.0 \nonumber\\
\left.\frac{1}{\tN}\frac{d^2 {\tilde N}}{d\Delta\phi}\right|_{\rm ATLAS}&\sim&1.8 \nonumber\\
\end{eqnarray}

There are additional subtleties:
\begin{itemize}
\item The CMS experiment presents the per-trigger-yield differential in $\Delta\phi$ between 0 to $\pi$.  There is a corresponding yield from 0 to $-\pi$ which is not shown nor included in the contribution of the associated yield (AY).  Both ALICE and ATLAS include the contribution from 0 to $-\pi$ when computing the AY thus making their results a factor of two larger.  When the ATLAS plots are shown from 0 to $\pi$ they fold over the results from $-\pi$ to 0 making the differential yield a factor of two larger.  ALICE plots the differential yields over the full phase space of $-\pi$ to $\pi$ therefore making a correction unnecessary. 
\item ALICE has a cut on their $\pp$ integrals such that $\ptrig \leq \ptasc$ for symmetric windows the signal will be a factor of two smaller than for asymmetric windows.
\end{itemize}

With these considerations, we obtain 
\begin{eqnarray}
\left.\frac{1}{\tN}\frac{d^2 {\tilde N}}{d\Delta\phi}\right|_{\rm CMS}\sim1.0\;\;&;&\;\;\left.\frac{1}{\Ntrig}\frac{d^2{\tilde N}}{d\Delta\phi}\right|_{\rm ATLAS}\sim 3.6 \nonumber \\
\left.\frac{1}{\tN}\frac{d^2 {\tilde N}}{d\Delta\phi}\right|_{\rm ALICE}^{\tiny{\rm{asym}}}\sim1.0 \;\;&;&\;\;
\left.\frac{1}{\tN}\frac{d^2 {\tilde N}}{d\Delta\phi}\right|_{\rm ALICE}^{\tiny{\rm{sym}}}\sim 0.5 \, .
\end{eqnarray}
and 
\begin{eqnarray}
\left.{\rm AY}\right|_{\rm CMS}\sim1.0\;\; &;&\;\; \left.{\rm AY}\right|_{\rm ATLAS} \sim 3.6  \nonumber\\
\left.{\rm AY}\right|_{\rm ALICE}^{\tiny{\rm{asym}}}\sim 2.0 \;\; &;&\;\;
\left.{\rm AY}\right|_{\rm ALICE}^{\tiny{\rm{sym}}}\sim 1.0 \, .
\end{eqnarray}

\bibliography{biblio}{}

\end{document}